\documentstyle[12pt,aaspp4]{article}
\begin{document}
\def\etal{et al.\ }
\def\vmi{\hbox{\it V--I\/}}
\def\kms{km s$^{-1}$}
\def\nd{\nodata}
\def\an{\AA}
\def\ll{$\lambda$}
\def\ma{$^{-1}$~}
\def\mb{$^{-2}$~}
\def\ha{H$\alpha$}
\def\deg{^{\circ}}
\def\Deg{\hbox{${}^\circ$\llap{.}}}
\def\Min{\hbox{${}^{\prime}$\llap{.}}}
\def\Sec{\hbox{${}^{\prime\prime}$\llap{.}}}
\def\hr{\hbox{${}^{\hbox{\sevenrm h}}$}}
\def\mn{\hbox{${}^{\hbox{\sevenrm m}}$}}
\def\deg{\hbox{${}^\circ$}}
\def\min{\hbox{${}^{\prime}$}}
\def\sec{\hbox{${}^{\prime\prime}$}}

\title {Photometric Recovery of Crowded Stellar Fields Observed with
{\it HST}/WFPC2 and the Effects of Confusion Noise on the
Extragalactic Distance Scale}

\author {Laura Ferrarese}
\affil{University of California Los Angeles, CA, 90095, USA, laura@astro.ucla.edu}
\authoraddr{8371 Math Sciences Building, Box 951562, Los Angeles CA 
90095-1562, USA}
\author {N.A. Silbermann}
\affil{Infrared Processing and Analysis Center, Caltech, Pasadena CA 91125, USA, nancys@ipac.caltech.edu}
\author {Jeremy R. Mould}
\affil{Research School of Astronomy \& Astrophysics, 
Institute of Advanced Studies, ANU, ACT 2611, Australia, jrm@mso.anu.edu.au}
\author {Peter B. Stetson}
\affil{Dominion Astrophysical Observatory, Victoria, British Columbia
V8X 4M6, Canada, Peter.Stetson@hia.nrc.ca}
\author {Abhijit Saha}
\affil{Kitt Peak National Observatory, NOAO, Tucson AZ 85726, USA, saha@noao.edu}
\author {Wendy L. Freedman}
\affil{Observatories of the Carnegie Institution of Washington, Pasadena, CA 91101, USA, wendy@ociw.edu}
\author {Robert C. Kennicutt, Jr.}
\affil{Steward Observatory, University of Arizona, Tucson, AZ 85721, USA, robk@as.arizona.edu}

\vskip .2in

\centerline{\it To appear in the February 2000 issue of the PASP}

\begin{abstract}

We explore the limits of photometric reductions of crowded stellar
fields observed with the Wide Field and Planetary Camera 2 on board
the {\it Hubble Space Telescope}. Two photometric procedures, based on
the DoPHOT and DAOPHOT/ALLFRAME programs are tested, and the effects
of crowding, complex sky background and cosmic-ray (CR) contamination
are discussed using an extensive set of artificial star simulations.
As a specific application of the results presented in this paper, we
assess the magnitude of photometric biases on programs aimed at
finding Cepheids and determining distances. We find that while the
photometry in individual images can be biased too bright by up to 0.2
mag in the most crowded fields due to confusion noise, the effects on
distance measurements based on Cepheid variables are insignificant,
less than 0.02 mag (1\% in distance) even in the most problematic
cases. This result, which is at odds with claims recently surfaced in
the literature, is due to the strict criteria applied in the selection
of the variable stars, and the photometric cross checks made possible
by the availability of multiple exposures in different filters which
characterizes Cepheid observations.

\end{abstract}


\section {INTRODUCTION}

During its six years of operation, the Wide Field and Planetary Camera
2 (WFPC2) on board the {\it Hubble Space Telescope} ({\it HST}) has
produced several thousand hours' worth of observations of crowded
stellar fields, from Galactic globular clusters to distant galaxies.
The photometric reduction of such fields is challenging. Much effort
has been spent in calibrating the photometric zero points and
assessing the magnitude of non-linearity effects (Holtzmann \etal
1995, Whitmore \& Hayer 1997, Hill \etal 1998, Stetson 1998, Saha
\etal 2000), but not enough attention has been devoted to
understanding whether and to what extent biases are introduced by the
procedures used for the photometric analysis.   Recently, Mochejska
\etal (1999) and Stanek \& Udalski (1999) addressed the effects of
confusion noise on distance measurements based on Cepheid variable
stars, and concluded that for galaxies at 20 Mpc observed with {\it
HST}, distances are underestimated by up to 15\% due to
blending. These studies are based on a simple extrapolation to large
distances of ground-based Cepheid observations in M31 and LMC
fields. Unfortunately,  they make two invalid assumptions: that the
photometric reduction procedures are incapable of recognizing and
correcting for the increasing level of crowding at larger distances,
and that the stellar background in the M31 and LMC fields are
representative of those of the more distant galaxies.   A more direct
assessment of the effects of crowding on the DoPHOT (Schechter \etal
1993) photometry of one particular galaxy, NGC 4639, which was
targeted for Cepheids by Saha \etal (1997) using {\it HST}, is
presented by Saha \etal (2000). For this galaxy, which is at a
distance of $\sim 25$ Mpc, the authors find a 5\% bias in distance due
to crowding, much smaller than inferred by Mochejska \etal and Stanek
\& Udalski, but still significant.

In this paper we test both DoPHOT and DAOPHOT/ALLFRAME (ALLFRAME from
now on for simplicity, Stetson 1994) photometric procedures using an
extensive set of artificial star simulations of fields spanning a wide
range of crowding and complexity of the underlying background. The
analysis presented in this paper has specific applications to the
results of the `{\it HST} Key Project on the Extragalactic Distance
Scale' (hereafter Key Project, Kennicutt, Freedman, \& Mould 1995),
and the `Type Ia Supernovae Calibration Project' (e.g., Sandage et
al. 1992), but is also relevant to any DoPHOT and ALLFRAME photometric
reduction of crowded stellar fields observed with the WFPC2.

The construction of the artificial star frames is discussed in \S 2,
and the procedure used in the photometric analysis in \S 3. The
results are presented in \S 4, \S 5 and \S 6. Section 4 describes
control tests done on uncrowded, constant background artificial star
fields, while \S 6 deals with the effects of confusion noise for both
DoPHOT and ALLFRAME. DoPHOT has also been subjected to two additional
test: Section 5 dissects the effects of CRs on the DoPHOT photometry
of crowded fields, and a simple test of the accuracy of the DoPHOT
photometric calibration is given in \S 7. These tests were done for
DoPHOT and not ALLFRAME without implying that we expect the
former to be more affected; similar tests could also be carried out
for ALLFRAME. Finally, \S 8 presents a summary of results.

\section {CONSTRUCTION OF THE ARTIFICIAL STAR FRAMES}

Of the fields targeted by the Key Project we selected two at opposite
extremes in terms of complexity of the background, confusion from
crowding and incidence of CR hits: the nearby galaxy NGC 2541
(d$\sim$12 Mpc) and NGC 1365 in the Fornax cluster (d$\sim$19 Mpc).
Most of the galaxies observed by the Key Project fall in between these
two extremes in terms of sky brightness and crowding, both across the
field and specifically in the regions where Cepheids are found, with
very few galaxies approaching the level of complexity presented by the
NGC 1365 field. Montages of the four WFPC2 CCDs for each field (from
which CRs have been removed) are shown in Figures 1 and 2. For each
galaxy, four out of the twelve existing epochs of observations were
selected, each with pairs of back to back exposures (referred to as a
`CR-split pairs') in both the F555W and F814W filters. Exposure times
and other relevant information are given in Table 1.

The artificial stars used to test both DoPHOT and ALLFRAME are created
using exclusively ALLFRAME parameters, in particular:  1) the average
PSFs constructed for the ALLFRAME analysis are assumed to be faithful
reproductions of the true system PSFs; and 2) the ALLFRAME photometric
calibration is used to transform the instrumental PSF magnitudes to
F555W and F814W magnitudes in the Holtzmann \etal (1995) ground based
system (see \S3).  In the ALLFRAME procedure followed by the KP, the
PSF is constructed from WFPC2 images of uncrowded stellar fields, and
is then used for the ALLFRAME photometric analysis of {\it all} of the
observed galaxies. However, the WFPC2 PSF changes with time due, for
example, to focus changes and telescope jitter:  these will lead to
systematic variations in the profile-fitting magnitudes in frames that
have true PSFs very different from the mean PSFs.  In the ALLFRAME
reduction stream, magnitude corrections are determined by measuring a
number of hand-selected local standards in each field  through a
series of synthetic apertures up to one-half arcsecond in radius.
These aperture magnitudes, projected to a fixed aperture radius of
one-half arcsecond through a growth-curve analysis (Stetson 1990) are
presumed to be insensitive to focus and tracking errors; the mean
difference between these aperture magnitudes and the profile-fitting
magnitudes for a given frame are applied to the latter as an `aperture
correction'.

DoPHOT takes the opposite approach: for each photometered frame, the
PSF is determined within the frame using suitable bright and isolated
stars throughout the field, so while different frames will generally
have different PSFs, the aperture corrections are maintained
relatively constant. Adding stars described by the ALLFRAME PSFs to an
existing galaxy frame has the inevitable  consequence  that the
artificial stars added to the images will by definition have the
fiducial PSFs and will not be subject to the same aperture corrections
as the actual stars in those same images. Therefore, the artificial
star experiments described here are not effective in testing the
validity of the aperture corrections or photometric zero points for
either reduction procedures: by definition, the artificial star
magnitudes recovered by ALLFRAME will agree {\it in the mean} with the
input magnitudes, while the errors in both the DoPHOT and ALLFRAME
aperture corrections will limit the significance of any difference
between the input and DoPHOT recovered magnitudes. Adopting an
ALLFRAME PSF for the artificial stars does not, however, affect our
ability to study crowding-induced photometric scatter and scale errors.

Two thousand artificial stars with given right ascension, declination
and $V$ and $I$ magnitudes were generated using the ALLFRAME
PSF and the luminosity functions shown by the solid black lines in
Figures 3 and 4. Figures 1 and 2 show images  of the artificial star
frames. Poisson noise was added to the artificial stars, however no
read noise or Poisson noise was added to the sky since these noise
sources are already present in the real galaxy frames. No artificial
stars were placed  close to the edges of the chips, which are
vignetted by the pyramid mirror.  The artificial stars were then added
to each of the NGC 2541 and NGC 1365 frames listed in Table 1.

\subsection{Nomenclature}
In what follows we will refer to the frames containing only the
artificial stars on a zero background as the `artificial frames', to
the original galaxy frames as the `real frames' and to the original
galaxy frames to which the artificial stars have been added as the
`real+artificial frames'. We will speak of an `exposure' in reference
to a single image of a CR-split pair (for example U2S72901T, see Table
1), and of an `epoch' in reference to the image obtained by combining
two CR-split exposures (for example U2S72901T and U2S72902T, see
\S3.1). `Input magnitudes' are the ones assumed in constructing the
artificial star frames, by definition they do not have an associated
error. `Recovered magnitudes' are the ones measured by running
ALLFRAME or DoPHOT on the frames; they will differ from the input
magnitudes due to the modeled photon statistics, the underlying read
noise and sky noise in the real frames, and the effects of crowding.

\section{PHOTOMETRIC PROCEDURE: APPLICATION OF ALLFRAME AND DoPHOT}

The real+artificial frames and the artificial frames were photometered
by NAS using  ALLFRAME (Stetson 1994) and by LF using a variant of
DoPHOT especially formulated to deal with the undersampling of WFPC2
data (Schechter, Mateo, \& Saha 1993; Saha \etal 1994). Neither NAS
nor  LF had access to the artificial star list (created by PBS) until
after the photometric analysis was complete. We followed the
methodology of the Key Project (e.g., Hill  \etal 1998,  Ferrarese
\etal 1996), except that only a subset of eight of the  24 Key Project
exposures of each field (four of 12 epochs), were used to derive the
master star list. As usual, all images are multiplied by four and
converted to integer before the photometric analysis is
performed. Therefore, the effective gain  is 1.75 e$^-$, as opposed to
7 e$^-$ of real WFPC2 data.

\subsection{DoPHOT Photometric Reduction}

Each pair of back to back CR-split exposures was combined -- to obtain
what we will be referring to as an `epoch frame'-- and CRs were
removed prior to the DoPHOT run. Further details of the DoPHOT
procedure can be found in Ferrarese \etal (1998) and Saha \etal
(1994).  Table 2 lists the PSF parameters (FWHM along the major and
minor axes and tilt angle) reported by DoPHOT for the real frames, the
artificial star frames, and the real+artificial frames. The artificial
stars' luminosity function is more heavily weighted towards bright
magnitudes, so while the PSF fitted by DoPHOT to the real frames is
slightly different from the artificial star PSF, the PSF fitted to the
real+artificial star frames is very close.

The DoPHOT aperture corrections needed to transform the fitted
magnitudes to the 0\Sec5 magnitudes defined by Holtzmann \etal (1995)
were calculated independently for each photometered frame, following
Ferrarese \etal (1998). Some of the DoPHOT aperture corrections are
listed in Table 3 for the artificial star frames, and for the first
exposure and first epoch (the two exposures of the first CR-split pair
combined, see Table 1) of the real+artificial star frames. The number
of stars used in deriving the aperture corrections is shown in
parentheses in each case. Because of jitter and focus changes,
aperture corrections for other exposures/epochs differ from those of
the first epoch and from each other by up to 0.05 mag. For comparison,
the aperture corrections derived for an uncrowded field in Leo I are
also shown. Notice that because the DoPHOT aperture corrections are
derived from a mixture of real and artificial stars, which (for
DoPHOT) are not described by an identical PSF, the
DoPHOT aperture corrections might not be appropriate for either  real
or artificial stars, and any zero point comparison quoted in this
paper is of very limited significance.  

Finally, the aperture corrected magnitudes were transformed to the
F555W and F814W magnitudes on the Holtzmann \etal ground based system
following Hill \etal (1998); the final transformation to $V$ and $I$
magnitudes, which requires the addition of a \vmi~ color term (the
same for DoPHOT and ALLFRAME, and up to 0.02 mag for F555W and 0.04
mag for F814W), is not necessary for our purpose.

\subsection{DAOPHOT/ALLFRAME Photometric Reduction}

As customary, ALLFRAME was run on all eight exposures
simultaneously. Notice that for a typical galaxy observed by the KP,
ALLFRAME would be run simultaneously on a larger  sample of $\sim 30$
frames, therefore in the case discussed here ALLFRAME has a less than
typical ability to recognize blended images.  The ALLFRAME fitted PSF
is, of course, identical to the artificial PSF by construction.  The
aperture corrections for the first and second exposures are as given
in Table 4, notice that these are more recent than the ones used by
Ferrarese \etal (1998) and Silbermann \etal (1999) for the original
reduction of the galaxy frames. As for the DoPHOT procedure, the
aperture corrected magnitudes are transformed to F555W and F814W
magnitudes  following Hill \etal (1998).

\section{CONTROL TESTS ON THE ARTIFICIAL FRAMES}

A control experiment was done by running DoPHOT and ALLFRAME on the
uncrowded, zero background, CR-free artificial star frames constructed
for NGC 1365 (the NGC 2541 artificial frames will not be discussed  as
they do not differ from the NGC 1365 frames in any relevant way). The
luminosity function of the recovered stars is shown in Figure 3 by the
thick red and green lines for ALLFRAME and DoPHOT.  Only the PC and
WF2 chips are shown as the results for the other two chips are
comparable. Notice that at faint magnitudes DoPHOT is more complete
than ALLFRAME in both chips, {\it a posteriori} we attribute this
result to  a more sensitive detection threshold (the number of sigma
above the background needed to trigger a detection) adopted in the
DoPHOT run. Figures 5 and 6 show a comparison of input magnitudes and
the magnitudes recovered by ALLFRAME and DoPHOT for the PC and WF2
(the results for the other chips are similar). A weighted mean and
standard deviation $\sigma$ (where the weights are given by the
inverse of the DoPHOT reported errors) of the difference $\Delta m$
between input and recovered magnitudes is calculated by excluding all
points deviating by more than 2$\sigma$ from the mean, and iterating
the process until convergence. This typically excludes 25\% of the
stars in F555W and 35\% in F814W. Stars fainter than 28 mag are also
excluded. The results are listed in  Table 5, where we also report the
slope $\gamma=[\Delta m$/input mag] and standard deviation derived
from a least-squares fit to all data points. A non-zero $\gamma$ would
indicate the presence of non-linearities in the photometry: for
example, if the sky were systematically overestimated, fainter stars
would be affected more than bright ones, and $\gamma$ would be
negative. As expected, both DoPHOT and ALLFRAME perform well on these
simple frames. No non-linearity effects are seen. Notice that the
perfect agreement in magnitude between input and recovered magnitudes
for ALLFRAME is artificial and results from having adopted the
ALLFRAME photometric calibration for the artificial stars. In view of
the fact that ALLFRAME aperture corrections for NGC 1365 are based
only on a handful of stars (Table 4), and the differences between the
ALLFRAME and DoPHOT PSFs, especially for the PC (Table 2), the
agreement between input and recovered magnitudes for DoPHOT is also
good. The solid flaring curves in Figures 5 and 6 represent the
typical error reported by DoPHOT and ALLFRAME for stars at any given
magnitude. The percentage of stars with difference between input and
measured magnitude larger than three times the reported $\sigma$ is
about 0.6\% for DoPHOT and 0.7\% for ALLFRAME, which confirms the
reliability of the error estimates as given by both photometric
procedures.

\section{EFFECTS OF COSMIC RAY HITS: DoPHOT TESTS}

ALLFRAME and DoPHOT take very different approaches when dealing with
CR hits.  Because ALLFRAME runs simultaneously on all
available exposures, information on the position and magnitudes of
each object is carried from frame to frame and CRs are easily
flagged.  DoPHOT identifies CRs as differing significantly
from a stellar PSF. Objects in each frame are identified with the help
of a master star list created from a deep, CR-free image
obtained by combining all exposures.  DoPHOT parameters can be set so
that objects identified in a single exposure but not present in the
master star list will be classified as CRs unless their PSF is
virtually identical to a stellar PSF. This flags the majority of
CR hits, but DoPHOT can still err when CRs hit  the
centers of stars included the master star list.   For this reason,
DoPHOT is more reliable when applied to frames from which CRs
have been removed.  This is an easy task for the Key Project galaxies
since, with very few exceptions,  CR-split pairs are available for
each epoch.  The CR-split exposures are combined  before processing
with DoPHOT and CRs are identified when the  difference in
counts between corresponding pixels (after accounting  for a global
difference in the sky level) is larger than four times a local sigma
calculated from the combined effects of Poisson statistics and local
noise.

This procedure poses some questions.  Because of the severe
undersampling of the WFPC2, particular care must be taken to assure
that the peaks of bright stars are not erroneously identified as CR
hits while combining the CR-split pairs.  In addition, unidentified CR
events  could lead to an overestimate or an underestimate of the
stellar  magnitudes, depending on whether the hits fall on top of a
stellar PSF  or in the nearby region used to measure the sky
respectively.  We have conducted  the following test to assess the
effects of CR events on the DoPHOT photometry.  For the
real+artificial star frames of both NGC 2541 and NGC 1365 we ran
DoPHOT independently on the first and second exposure of the first
epoch for each galaxy, which are heavily affected  by CRs, and on the
CR-cleaned first epoch frame, produced by combining the first and
second exposure.  In all cases the master star list derived from all
exposures combined was used.  Figure 7 shows a comparison of sky
values measured in the first exposure and the first epoch for the
F555W filter and WF2 chip, which for both galaxies shows the largest
background excursions.  The total counts (DN) collected for the sky
are given, with the number of counts for the single exposure scaled to
match the exposure time of the combined frame.

Two conclusions can be drawn.  First, the mean sky difference  at low
background levels is not zero.  Inspection of the frames shows  that
this corresponds to a difference in the level of scattered light
between the two consecutive frames due to the change in orbital
position of the spacecraft.  More importantly, there is a
non-linear  effect.  A least-squares fit to all available data
points gives $\Delta  s / s = (0.0094\pm0.0004)$
for NGC 2541, and  $\Delta s /s = (0.0093\pm0.0002)$ 
for NGC 1365,  where $\Delta s$ is the difference in sky values
between the first exposure and the first epoch of the CR-split pair
(scaled to the same exposure time), and $s$ is the sky value for the
first epoch.  The sense of the difference is that the sky is measured
brighter in a single, CR-affected exposure, the more so the higher the
background level.  Notice that the agreement in the slopes measured
for the NGC 2541 and NGC 1365 frames (which have exposure times
differing by over a factor two) implies that the effect depends on
total number of counts in the sky rather than on sky surface
brightness.  Responsible for this scale error is the presence, in the
single exposures, of CR hits which are not flagged and therefore are
folded in when calculating the background.  Because hits are more
easily missed on brighter backgrounds, this produces a scale error in
the sky measurements.  In the epoch frame, the CR hits are correctly
identified and removed, and the sky is measured accurately. We
therefore recommend that CR-cleaned frames be used for DoPHOT
reductions whenever possible, as was done for the photometric
reduction of the galaxies observed as part of the KP and the Type Ia
Supernovae Calibration Project.

How does this affect the  stellar magnitudes?  Figure 8 shows the
magnitude difference $\Delta m$ between the first exposure and first
epoch of the WF2 CR-split pair: these are the stars whose background
is plotted in Figure 7.  Given the slope of the scale error measured
in the sky, and assuming that the stellar magnitudes are not affected
by the presence of CRs, two stars of 25th and 27th magnitude
projecting onto a sky background of 20 mag arcsec$^{-2}$
(corresponding to 200 and 480 DN in the bottom and top panel of Figure
7 respectively) would be measured too faint by 0.01 and 0.1 mag
respectively. This would produce a scale error when comparing
magnitudes recovered from frames with and without CRs.  No obvious
scale error is however observed in Figure 8: a least square fit to the
data points produces slopes $\gamma=[\Delta m$/input mag] smaller than
0.01 mag/mag. Several causes contribute to this result. First of all,
faint stars are more easily lost on brighter backgrounds. Figure 9
shows the recovery rate for stars of given magnitude as a function of
backgrounds brightness. Therefore, most of the stars that would show
the largest deviations, and contribute more to produce a scale error
(i.e. faint stars on bright backgrounds) are not even
detected. Second, just as the presence of CRs leads DoPHOT to
overestimate the background brightness, it is reasonable to expect
that the stars magnitudes will be somewhat overestimated as well, with
the consequence that the two effects cancel, at least
partially. Support to this point comes from the fact that when the
difference $\Delta m$ is plotted as a function of sky brightness, no
obvious correlation is observed.

Figure 8 shows an increased scatter in the sense that stars  are
measured preferentially brighter in the single exposure than in  the
CR-free epoch frame. Visual inspection of the images  confirms that
these (the red crosses in Figure 8) are unfortunate stellar images
that happen to be hit head on by a CR.  The CR  is not identified in
the single exposure, and the star is measured too  bright.
Unfortunately, these stars cannot be discriminated based on  their
reported error $\sigma$: for all of them $\Delta m > 4\sigma$ and
their reported photometric errors are perfectly normal for their
measured magnitude.  Therefore, the bias  would be undetected  if no
combined, CR-free images were available.

Is the problem  solved in the case of the Key Project galaxies, for
which DoPHOT is run on CR-cleaned images?  Not entirely.  Figure  10
shows the same red stars as in Figure 8 but this time the values
plotted are the magnitude difference between the first exposure and
the  first epoch (in red as in Figure 8),  between the second exposure
and the first epoch (in black), and  between the second exposure and
the deep frame  created by combining all available exposures (in
blue).  It can be readily seen that the blue points have a mean
positive $\Delta m$,  while the black points are well distributed
around a zero $\Delta m$.   We remind the reader that the stars
plotted here were selected by the circumstance of having been affected
by a CR in the first exposure; in very few cases will the star also
have been affected by a CR in the second exposure.  The fact that the
black points show systematically positive magnitude residuals (by
0.10$\pm$0.16 mag) argues that the procedure used in combining the
first and second exposures to produce the epoch frame must leave some
of the CR hit behind, so that the star is still measured brighter in
the combined frame than in the second exposure. However, when all
eight available frames are combined to create the deep image, the
process of eliminating cosmic rays is much more effective, and the
bias disappears (blue points; the weighted mean in this case is
0.01$\pm$0.17).  In conclusion, the simulations show that when DoPHOT
is applied following the KP prescriptions, about 3\% of the stars at
any given epoch are measured too bright by an average of 0.1 mag.  For
the KP galaxies, which are observed for a total of 12 epochs, this
means that about 1/3 of the Cepheids will have one out of twelve
measurements biased high by 0.1 mag, which has no impact on the
period-luminosity (PL) relation.

\section{EFFECTS OF CONFUSION: TESTS ON THE REAL+ARTIFICIAL FRAMES}

We now turn our attention to the NGC 1365 and NGC 2541 frames to which
the artificial stars have been added. From now on,   DoPHOT is always
applied to frames which are CR-cleaned (\S 3), while ALLFRAME uses the
original, CR-split exposures.  The luminosity function of the recovered
artificial stars is shown in Figures 3 and 4 by the thin yellow and
blue lines for the PC and WF2 chips and the ALLFRAME and DoPHOT
reductions.  Figures 11 and 12 show the comparison between input and
recovered magnitudes  for both ALLFRAME and DoPHOT. The results do not
vary significantly from chip to chip for the same filter, and we only
plot them for the WF2 chip of both NGC 1365 and NGC 2541. The same
parameters listed in Table 5 for the artificial frames, i.e., the
weighted mean difference between input and recovered magnitudes, and
the slope of a least-squares fit line through the data, can be found
in Table 6 for the real+artificial frames. Again, no significant scale
errors are found, and the agreement in the zero points is within the
uncertainties in the DoPHOT and ALLFRAME aperture corrections (see \S
3). Figures 11 and 12, however, deserve some closer inspection. The
distributions show larger scatter for NGC 1365 than for NGC 2541, as
expected due to the more crowded nature of the former. Unlike the
control case described in \S 4, however,  the number of points which
deviate by more than three times the reported error does not follow a
Gaussian distribution. In NGC 1365, 16\% and 24\% of the F555W and
F814W DoPHOT stars brighter than 28 mag deviate by more than
$3\sigma$, in NGC 2541 the percentages are 14\% and 21\%.  The same
result, but not quite as extreme has been discussed by Saha \etal
(2000), who further point out that this translates into a bias of up
to 0.1 mag in the Cepheid distance moduli derived using DoPHOT in
crowded fields.  The plots based on ALLFRAME photometry have larger
scatter, however, the photometric errors reported by ALLFRAME are
larger than measured  by DoPHOT, with the consequence that a smaller
fraction of the ALLFRAME stars, 8\% and 5\% in NGC 1365 and NGC 2541
respectively, deviate by more than $3\sigma$.

For all stars deviating by more than $3\sigma$ (the circles in Figures
11 and 12) we plot in Figure 13 the difference between the DoPHOT
magnitudes measured in the first epoch and in all subsequent
epochs. Because the main features of this plot repeat for all chips
and both filters, we only show the case of the WF2 F814W reduction in
NGC 1365.  Some of the points (shown by the circles) are highly
correlated.  The magnitudes for these stars are mis-measured in the
first epoch because of some transient event, such as the presence of a
nearby hot pixel, or an undetected CR, but are measured correctly in
all other epochs. While there is really no way to identify these stars
in single epoch programs, in  multi-epoch observations (such as 
Cepheid finding programs)  the corrupted epoch would easily be
flagged.  The points shown by the crosses, however, show no obvious
correlation: they scatter around a zero mean (within the uncertainties
in the aperture corrections) and show no scale errors. These stars are
measured systematically too bright in all of the epochs. The reason is
confusion due to unresolved stellar blends or rapidly changing
background level with position on the chip. Because these patterns
repeat identically from epoch to epoch, they produce a bias which is
impossible to detect unless artificial star experiments, of the type
described in this paper, are performed in each individual case.
Following Saha \etal (2000) we plot in Figures 14 to 15 the
correlations between the F555W and F814W residuals for the first epoch
of both galaxies, and for both ALLFRAME and DoPHOT. The results shown
for the PC and WF2 are representative of the other chips. In each
figure crosses are for stars that deviate by more than $3\sigma$ in
both passbands, filled and open circles for stars that deviate by more
than 3$\sigma$ in either F555W or F814W, and  dots for well behaved
stars in both bands. In all cases, we notice that there is no
correlation for stars which are not affected by confusion noise (dots
and circles). In Figure 13, these stars would be plotted as circles,
i.e., they happen to be measured incorrectly in one particular epoch
and one particular filter due to some reason that does not repeat for
all other epochs (for example the vicinity of a CRs hit, or defecting
pixels), and would therefore not produce a bias in the derivation of
Cepheid distance moduli. However, there is a strong correlation for
stars for which the difference between measured and expected magnitude
is statistically larger {\it in both passbands} than expected given
their reported photometric errors (crosses). As pointed out above, the
magnitude of these stars is overestimated in each epoch and both
filters, and will introduce a bias in the derived Cepheid PL
relations. A quantitative discussion of the photometric biases arising
from the effects described above when the photometry is performed on
single frames is discussed in \S6.1. Section 6.2 applies to the
specific case of the Cepheid observations carried out by the KP.

\subsection {Photometric Bias in Single Images}

The amount of bias from confusion noise in single-epoch observations
can be quantified from plots such as the ones shown in Figure 11 and
12:  the bias is simply represented by how much, in the mean, the
recovered magnitudes differ from the input magnitudes. This is
estimated by:

$\bullet$  defining a robust weighted mean $\overline{\Delta
m}_{true}$ of the difference between input and recovered magnitudes in
the real+artificial frames as described in \S 4. Only bright,
isolated, and therefore well measured stars contribute to
$\overline{\Delta m}_{true}$, which therefore represents the `true'
value of the mean for any sample of stars in the absence of biases;

$\bullet$ computing a straight mean $\overline{\Delta m}_{bias}$ of
all available data points recovered in the real+artificial frames;

$\bullet$ calculating the difference  $\epsilon = \overline{\Delta
m}_{true} - \overline{\Delta m}_{bias}$. This represents the amount of
bias for single epoch observations,  and is listed in Table 7.

We note that while for single epoch one band photometry of crowded
fields ALLFRAME is more affected by confusion noise than DoPHOT (Table
7), ALLFRAME was specifically  designed to take optimum advantage of
the information content of multi-filter, multi-epoch observations,
such as searches for variable stars. While ALLFRAME shows larger
scatter than DoPHOT in Figures 11 and 12, the deviant points have
large photometric errors in ALLFRAME, which is not the case for
DoPHOT. While it would be difficult to discriminate against these
stars in individual exposures (apart from rejecting them on the basis
of their large errorbars), the problem is lessened in most practical
cases: photometric observations often involve multiple filters,
for example, and for Cepheids observations carried out using the KP
procedure, the large photometric errors attached to the ALLFRAME
magnitudes would prevent one from selecting most of the deviant stars
as Cepheid variables.

\subsection {Photometric Bias in the Extragalactic Distance Scale}

For multi-epoch programs such as observations of Cepheid variables,
the bias discussed in \S 6.1 is somewhat lessened. This is because
some of the deviant stars will be flagged once the photometry obtained
for consecutive exposure or in different filters is compared.  Let us
consider the case of the galaxies observed by the KP, whose goal is to
discover Cepheids and measure their period and mean
magnitudes. Observation in two filters and multiple epochs are
available. First of all, we pointed out already that stars which
deviate in {\it only one} of the filters by more than three times
their reported error (Figures 14 and 15) are corrupted by a transient
phenomenon (for example a cosmic ray) affecting that particular epoch,
but none of the others. When comparing photometry from different
epochs, the corrupted measurement would easily be flagged (this would
correspond, for example, to an obviously deviant point in an otherwise
well phased light curve). We can therefore remove the corrupted
measurements from our sample.  Second, fits to the Cepheid PL relation
are not weighted, however Cepheids which deviate from the ridge-line
of the PL relation by a significant amount (generally three times the
intrinsic 1$\sigma$ width of the relation) are rejected. The KP
adopted $1\sigma=0.27$ mag in $V$ and $1\sigma=0.18$ mag in $I$ for
the PL relation. When calculating the bias introduced by confusion
noise in the Cepheid sample we do the following (the artificial star
magnitudes in this case can be related to the Cepheids mean
magnitudes):

$\bullet$ include in the sample only measurements that deviate from
the straight mean of all data points by more {\it or} less than three
times their reported error in {\it both} F555W or F814W. This
addresses the first point mentioned above. Farther reduce the sample
by excluding all stars which deviate from the mean by more than 0.81
mag in F555W and more than 0.54 mag in F814W. This addresses the
second point above and flags only a few of the most extreme points.

$\bullet$ For this sample, calculate $\overline{\Delta m}_{true}$ as
done in the case of single-epoch observations. Effectively, because
$\overline{\Delta m}_{true}$ is a robust weighted mean, the prior
exclusion of the data satisfying the conditions above
has no effects.

$\bullet$ For the same sample, calculate the mean difference
$\overline{\Delta m}_{bias}$ between input and recovered magnitudes.
Given the distances to NGC 2541 and NGC 1365, and the fact that the
maximum period (and therefore maximum mean magnitude) of the Cepheids
is constrained by the length of the observing window (~60 days),  the
range in $V$ magnitudes spanned by the Cepheids is $24.3 < m_V << 28$
mag and $25.1 < m_V << 28$ mag in  NGC 2541 and NGC 1365 respectively
(Ferrarese \etal 1998, Silbermann \etal 1999).  Only these magnitude
ranges are therefore considered in calculating $\overline{\Delta
m}_{bias}$.  Notice that the exclusion of the brighter, least biased,
part of the distribution in Figures 11 and 12, and the inclusion of
the faint tail up to $m_V = 28$ mag (when in fact few Cepheids ever
get fainter than 27.5 mag) leads to an increase in $\overline{\Delta
m}_{bias}$, compared to the case in which the entire wavelength range
were to be used.

$\bullet$ Finally, calculate the difference $\epsilon =
\overline{\Delta m}_{true} - \overline{\Delta m}_{bias}$. This is the
bias due to confusion noise expected in the {\it apparent} Cepheid
distance moduli, and is listed in Table 8.

The bias in the {\it final}, dereddened distance modulus is given by
$\delta \mu_0 \sim \epsilon_V$ when the magnitudes in both $V$ and $I$
passbands are mis-measured by approximately the same amount (i.e. the
correlations of residuals in Figures 14 and 15 has a unit slope) (Saha
\etal 2000). From Table 8, the first conclusion is that even for the
most crowded fields, i.e. WF2 of NGC 1365, the bias is not larger than
0.07-0.08 mag, in agreement with the findings by Saha \etal (2000) in
the very crowded field of NGC 4639 (the authors estimate the bias to
be less than 0.08 mag). In the specific example of NGC 1365, for which
Silbermann \etal (1999) derived a distance modulus of $31.31 \pm$ 0.20
(random) $\pm$ 0.18 (systematic) based on ALLFRAME photometry, the
bias estimated from a mean of all four chips in Table 8 would be $\sim
0.025$ mag, and can be easily neglected. In the less crowded fields of
more nearby galaxies, such as NGC 2541, the bias is further reduced.

There is an additional factor which, while not easily quantifiable,
will farther decrease the already insignificant amount of bias in
the Cepheid distance modulus.  Contamination by an underlying
companion would artificially decrease the amplitude of the Cepheid
light curve, and either lead to a reduction of confidence in the light
curve, or preclude detection as a variable star altogether. We have
conducted experiments by adding increasing amount of
contamination to real Cepheid light curves observed by the Key Project
Team. While we could not discriminate against Cepheids for which the
contamination amounts to  30\% or less of the Cepheid mean flux, in
most cases the Cepheids were recognized as affected when the
contamination level was increased beyond 60\%.

One last points needs to be discussed. The bias calculated above can
be  considered representative of the bias affecting the Cepheid
sample, and therefore the Cepheid distance modulus, only if the
artificial stars and the Cepheids are affected by the same level of
crowding. Figure 16 shows the mean and rms contamination affecting the
sample of Cepheids (solid circles and smaller errorbars) and
artificial stars (open circles and larger errorbars) as a function of
F555W magnitudes. To calculate the quantity on the y-axis, we counted
the number of real stars detected within a 5 pixel radius of each
Cepheid and artificial star. We then calculated the ratio between the
total flux contributed by these nearby companions to the flux of the
central star. We refer to this quantity, multiplied by 100, as the
`percent contamination': its mean and standard deviation,  binned in
0.5 magnitude intervals, are plotted as a function of the F555W
magnitude of the central star in Figure 16, for both NGC 2541 and NGC
1365. In making the plot, we used the real sample of Cepheids from
which a distance was derived (from Ferrarese \etal 1998 and Silbermann
\etal 1999).  So, for example, the total flux of the stars within a 5
pixel radius of the NGC 2541 Cepheids with F555W magnitudes between
24.5 and 25.0 is equal, in the mean, to $f_c/f = (4\pm5)$\% the flux
of the Cepheid. In the same magnitude range, $f_c/f = (5\pm12)$\% for
the artificial stars. It is easy to see from the figure that, if
anything, the artificial stars are {\it more} contaminated than the
Cepheids.  Therefore the bias listed in table 8 can be considered as a
hard upper limit to the amount of bias affecting the Cepheid sample,
and the derived Cepheid distances.

The results of the tests presented in this work are at odds with the
finding of Mochejska \etal (1999) and Stanek \& Udalski (1999), who
speculate from the analysis of Cepheids in LMC and M31 fields that the
Cepheid distances published by the KP and the Type Ia Supernovae
Calibration Team are underestimated by up to 15\% (0.3 mag)  due to
neglecting the effects of blending.  The Mochejska  \etal and Stanek
\& Udalski analysis, while correctly assuming that unresolved blending
will artificially increase a star's magnitude, do not consider that
for the same reason the underlying sky will be brightened. The
interplay of the two effects cannot be estimated unless the photometry
is actually {\it performed on} (as opposed to extrapolated to) the
distant galaxy fields, using some kind of control test which in our
case is provided by the artificial stars. In addition,  Mochejska
\etal and Stanek \& Udalski assume that the stellar background in the
LMC and M31 fields is  representative of the more distant fields
observed with HST, when in fact it is significantly brighter.

\section{PHOTOMETRIC ZERO POINT TEST: DoPHOT}

The tests discussed so far say very little about the accuracy of the
DoPHOT or ALLFRAME zero points, since the artificial star magnitudes
are based on the ALLFRAME photometric calibration. The DoPHOT zero
point was tested for the most crowded of the fields considered, the
WF2 chip of NGC 1365, using as `artificial stars' a real field
observed in the dwarf local group galaxy Sextans A. The Sextans A
images used were obtained as part of program GO-5915 on December 1,
1995, for a total of 1800 seconds of integration in both F555W and
F814W filters. Based on \S 4, we expect the magnitudes reported by
DoPHOT when run on this  uncrowded field to be very accurate. A
comparison between these magnitudes, and the magnitudes recovered when
the Sextans A field is added to the F555W WF2 NGC 1365 frame is shown
in Figure 17. A weighted fit to all stars gives perfect agreement in
the absolute zero points, $\Delta(m) = 0.000\pm0.065$, and no scale
error.

\section{SUMMARY OF CONCLUSIONS}

We have discussed the performance of the DoPHOT and ALLFRAME
photometric procedures when applied to crowded fields observed with
the WFPC2 on board {\it HST}. We have focused on searching for biases
that can affect the determination of the Cepheid distance moduli
derived by the {\it HST} Key Project on the Extragalactic Distance
Scale and the Type Ia Supernovae Calibration Project, but our results
are relevant for any photometric study requiring high precision
measurements.  The following conclusions and guidelines are the
results of this paper.

$\bullet$ When doing any photometric analysis, tests should be
performed to asses the impact of CR events on the photometry. In the
case of DoPHOT,  it is highly recommended to use frames from which CRs
have been removed. When CRs are present, DoPHOT has a tendency to
overestimate  the sky brightness by amounts that can be significant
when total counts per pixel in the sky exceed a few hundred DN. The
problem is completely solved only when several exposures are available
and the photometry is obtained from an image created by combining all
exposures. In the specific case of most of the galaxies observed as
part of the {\it HST} Key Project, where only two exposures are
available for each epoch of observation, the stars affected are
different in each epoch and are not present in a large enough number
to introduce a bias in the derived Cepheid distance modulus. The
effects of CRs on the ALLFRAME photometry remain to be tested, however
given the good agreement between the ALLFRAME and DoPHOT photometry
obtained for all KP galaxies (better than 0.05 mag, or well within the
uncertainties in the aperture corrections), it is unlikely that CRs
introduce a significant bias, if any.

$\bullet$ Both DoPHOT and ALLFRAME perform extremely well, both in 
recovering magnitudes and in estimating their errors, in uncrowded
fields with limited variation in the background brightness across the
chip.

$\bullet$ In crowded fields, such as the ones observed by the
{\it HST} Key Project team, a bias is introduced by the presence of
confusion noise due to crowding and rapidly varying background levels
across the chip. This is in the sense that a  fraction ($\sim$ 5-10\%
for ALLFRAME and up to 25\% for DoPHOT) of the stars are measured
consistently too bright. The effect on the photometry of single epoch
observations can be significant, 0.05 mag even for not too crowded
fields, such as the WF4 chip of NGC 2541, and up to 0.2 mag for the
most crowded and distant fields observed by the KP, such as NGC
1365. The bias is  somewhat worse when ALLFRAME, rather than DoPHOT
photometry is used.  When the photometry is applied to derive Cepheid
distance moduli from multi-epoch observations, however, the bias is
significantly reduced because strict criteria are imposed in selecting
the variable stars. In the case of ALLFRAME, the bias in the final
distance modulus is negligible, leading to underestimate the distance
by 1\% (0.02 mag) at most for the most crowded and distant fields
observed by the {\it HST} Key Project. The effects is slightly larger,
$\sim$ 2\% mag, when DoPHOT photometry is used, but is in no case as
large as extrapolated by Mochejska \etal (1999) and Stanek \& Udalski
(1999) from ground based studies of M31 and LMC fields.

\acknowledgments

We wish to thank the anonymous referee for the useful suggestions that
helped improve the quality of this manuscript.  LF acknowledges
support by NASA through Hubble Fellowship grant HF-01081.01-96A and
through Long Term Space Astrophysics program NRA-98-03-LTSA-03.

\appendix

\section{An Empirical Test of the Effects of Crowding on the Distance Scale}

We have concluded that the Key Project distance scale is compressed by
photometric confusion problems by no more than 0.025 mag. It is
interesting to consider whether there is any way of checking that
$a~posteriori$.

One possibility is to examine residuals from the Tully Fisher relation
and to seek correlations with distance. If distances for remote
galaxies are underestimated due to their higher probability of image
blending (Stanek \& Udalski 1999), those galaxies would lie  low in
the Tully Fisher diagram of Sakai \etal (2000), as also pointed out by
Gibson \etal (2000).

Figure 18 shows H-band residuals from Sakai's equation (10) plotted
against distance.  Such correlation as there is is in the opposite
sense. In fact, calculating the slope in Figure 18 and its
uncertainty, we can rule out the $positive$ correlation one might
expect due to blending at the 1.85$\sigma$ level.  This supports our
conclusion that the Key Project distance scale is  not compressed by
photometric confusion problems by more than 0.025 mag, a conclusion
also reached by Gibson \etal


\begin{deluxetable}{llrrr}
\tablecolumns{5} \tablewidth{0pc} \scriptsize \tablecaption{Data
Files\label{tbl-1}} \tablehead{ \colhead{Description\tablenotemark{1}}
& \colhead{Rootname} & \colhead{t$_{exp}$(s)} & \colhead{Filter} &
\colhead{Date Obs.} \\ } \startdata \multicolumn{4}{l} {--NGC
2541--}\nl 1st V epoch, 1st exposure &U2S72901T & 1100 & F555W &
30/10/95\nl 1st V epoch, 2nd exposure &U2S72902T & 1100 & F555W &
30/10/95\nl 1st I epoch, 1st exposure &U2S72903T & 1300 & F814W &
30/10/95\nl 1st I epoch, 2nd exposure &U2S72904T & 1300 & F814W &
30/10/95\nl 2nd V epoch, 1st exposure &U2S73001T & 1100 & F555W &
5/11/95\nl 2nd V epoch, 2nd exposure &U2S73002T & 1100 & F555W &
5/11/95\nl 2nd I epoch, 1st exposure &U2S73003T & 1300 & F814W &
5/11/95\nl 2nd I epoch, 2nd exposure &U2S73004T & 1300 & F814W &
5/11/95\nl 3rd V epoch, 1st exposure &U2S73402T & 900  & F555W &
20/11/95\nl 3rd V epoch, 2nd exposure &U2S73403T & 900  & F555W &
20/11/95\nl 3rd I epoch, 1st exposure &U2S73405T & 1100 & F814W &
20/11/95\nl 3rd I epoch, 2nd exposure &U2S73406T & 1100 & F814W &
20/11/95\nl 4th V epoch, 1st exposure &U2S73901T & 1100 & F555W &
8/12/95\nl 4th V epoch, 2nd exposure &U2S73902T & 1100 & F555W &
8/12/95\nl 4th I epoch, 1st exposure &U2S73903T & 1300 & F814W &
8/12/95\nl 4th I epoch, 2nd exposure &U2S73904T & 1300 & F814W &
8/12/95\nl \multicolumn{4}{l} {--NGC 1365--}\nl 1st V epoch, 1st
exposure &U2S71201T & 2400 & F555W & 19/09/95\nl 1st V epoch, 2nd
exposure &U2S71202T & 2700 & F555W & 19/09/95\nl 1st I epoch, 1st
exposure &U2S71203T & 2700 & F814W & 19/09/95\nl 1st I epoch, 2nd
exposure &U2S71204T & 2700 & F814W & 19/09/95\nl 2nd V epoch, 1st
exposure &U2S70201T & 2400 & F555W &  6/08/95\nl 2nd V epoch, 2nd
exposure &U2S70202T & 2700 & F555W &  6/08/95\nl 2nd I epoch, 1st
exposure &U2S70203T & 2700 & F814W &  6/08/95\nl 2nd I epoch, 2nd
exposure &U2S70204T & 2700 & F814W &  6/08/95\nl 3rd V epoch, 1st
exposure &U2S70301T & 2400 & F555W & 14/08/95\nl 3rd V epoch, 2nd
exposure &U2S70302T & 2700 & F555W & 14/08/95\nl 3rd I epoch, 1st
exposure &U2S70303T & 2700 & F814W & 14/08/95\nl 3rd I epoch, 2nd
exposure &U2S70304T & 2700 & F814W & 14/08/95\nl 4th V epoch, 1st
exposure &U2S70701T & 2400 & F555W & 29/08/95\nl 4th V epoch, 2nd
exposure &U2S70703T & 2300 & F555W & 29/08/95\nl 4th I epoch, 1st
exposure &U2S70705T & 2300 & F814W & 29/08/95\nl 4th I epoch, 2nd
exposure &U2S70706T & 2700 & F814W & 29/08/95\nl \enddata
\tablenotetext{1}{The complete time sequence observed by the KP for
NGC 2541 and NGC 1365 comprises a total of 12 epochs. The ones
considered here have been labeled first to fourth for convenience.}
\end{deluxetable}

\begin{deluxetable}{lrrrrrrrrrrrr}
\tablecolumns{13}
\tablewidth{0pc}
\scriptsize
\tablecaption{DoPHOT PSF Parameters\label{tbl-1}}
\tablehead{
\colhead{} &
\multicolumn{3}{c}{PC} &
\multicolumn{3}{c}{WF2} &
\multicolumn{3}{c}{WF3} &
\multicolumn{3}{c}{WF4}\\
\colhead{} &
\colhead{$\Delta_{maj}$} &
\colhead{$\Delta_{min}$} &
\colhead{Tilt} &
\colhead{$\Delta_{maj}$} &
\colhead{$\Delta_{min}$} &
\colhead{Tilt} &
\colhead{$\Delta_{maj}$} &
\colhead{$\Delta_{min}$} &
\colhead{Tilt} &
\colhead{$\Delta_{maj}$} &
\colhead{$\Delta_{min}$} &
\colhead{Tilt} \\
\colhead{} &
\colhead{(pix)} &
\colhead{(pix)} &
\colhead{} &
\colhead{(pix)} &
\colhead{(pix)} &
\colhead{} 
&\colhead{(pix)} &
\colhead{(pix)} &
\colhead{} 
&\colhead{(pix)} &
\colhead{(pix)} &
\colhead{} \\}
\startdata
\multicolumn{10}{l} {--NGC1365, F555W--}\nl
Real frame            & 1.382 & 1.318 &  30\Deg62 & 1.275 & 1.262 & 
$-$22\Deg80 & 1.272 & 1.226 &  21\Deg16 & 1.260 & 1.140 & $-$68\Deg44 
\nl
Artificial frame      & 1.271 & 1.109 &  37\Deg71 & 1.174 & 1.130 & 
$-$20\Deg81 & 1.328 & 1.224 &  37\Deg23 & 1.299 & 1.159 & 
$-$67\Deg72  \nl
Real+art., 1st exp.   & 1.257 & 1.138 &  32\Deg97 & 1.168 & 1.133 & 
$-$42\Deg76 & \nd   & \nd   & \nd    & \nd   & \nd   & \nd     \nl
Real+art., 1st epoch  & 1.274 & 1.163 &  33\Deg65 & 1.159 & 1.114 & 
$-$39\Deg15 & 1.263 & 1.165 &  31\Deg70 & 1.251 & 1.120 & $-$68\Deg35 
\nl
\multicolumn{10}{l} {--NGC1365, F814W--}\nl
Real frame            & 3.924 & 3.819 & 39\Deg14 & 1.291 & 1.226 &  
80\Deg99 & 1.299 & 1.223 &   1\Deg96 & 1.313 & 1.213 & $-$80\Deg03  
\nl
Artificial frame      & 3.987 & 3.837 & 43\Deg42 & 1.160 & 1.143 &  
75\Deg78 & 1.246 & 1.161 &  32\Deg03 & 1.240 & 1.153 & $-$72\Deg04  
\nl
Real+art., 1st exp.   & 3.810 & 3.734 & 42\Deg62 & 1.206 & 1.175 & 
$-$74\Deg43 & \nd   & \nd   & \nd    & \nd   & \nd   & \nd     \nl
Real+art., 1st epoch  & 3.856 & 3.761 & 48\Deg32 & 1.196 & 1.166 &  
88\Deg76 & 1.217 & 1.154 &  34\Deg38 & 1.223 & 1.144 & $-$76\Deg32 \nl
\multicolumn{10}{l} {--NGC2541, F555W--}\nl
Real frame            & 1.198 & 1.118 &  37\Deg35 & 1.195 & 1.112 & 
$-$36\Deg04 & 1.199 & 1.122 &  17\Deg37 & 1.226 & 1.074 & $-$70\Deg42 
\nl
Artificial frame      & 1.247 & 1.111 &  39\Deg05 & 1.171 & 1.123 & 
$-$30\Deg70 & 1.329 & 1.221 &  31\Deg99 & 1.285 & 1.151 & 
$-$64\Deg45  \nl
Real+art., 1st exp.   & 1.268 & 1.127 &  34\Deg42 & 1.149 & 1.113 & 
$-$20\Deg23 & \nd   & \nd   & \nd    & \nd   & \nd   & \nd     \nl
Real+art., 1st epoch  & 1.273 & 1.144 &  35\Deg21 & 1.145 & 1.113 & 
$-$26\Deg46 & 1.255 & 1.164 &  30\Deg92 & 1.201 & 1.092 & $-$63\Deg09 
\nl
\multicolumn{10}{l} {--NGC2541, F814W--}\nl
Real frame            & 3.764 & 3.651 &  42\Deg69 & 1.189 & 1.151 & 
$-$47\Deg64 & 1.185 & 1.148 &  13\Deg28 & 1.206 & 1.107 & $-$72\Deg35 
\nl
Artificial frame      & 4.010 & 3.848 &  44\Deg99 & 1.158 & 1.148 & 
$-$49\Deg09 & 1.227 & 1.169 &  27\Deg21 & 1.225 & 1.156 & $-$67\Deg85 
\nl
Real+art., 1st exp.   & 3.870 & 3.733 &  41\Deg85 & 1.152 & 1.131 &   
6\Deg67 & \nd   & \nd   & \nd    & \nd   & \nd   & \nd \nl
Real+art., 1st epoch  & 3.898 & 3.753 &  41\Deg58 & 1.163 & 1.156 & 
$-$85\Deg61 & 1.191 & 1.154 &  36\Deg68 & 1.218 & 1.135 & $-$76\Deg48 
\nl
\enddata
\end{deluxetable}

\begin{deluxetable}{lllll}
\tablecolumns{5}
\tablewidth{0pc}
\scriptsize
\tablecaption{DoPHOT Aperture Corrections\label{tbl-1}}
\tablehead{
\colhead{} &
\colhead{PC (mag)\tablenotemark{1}} &
\colhead{WF2 (mag)\tablenotemark{1}} &
\colhead{WF3 (mag)\tablenotemark{1}} &
\colhead{WF4 (mag)\tablenotemark{1}} \\
}
\startdata
\multicolumn{5}{l} {--F555W--}\nl
LeoI values                   & $-$0.880                  & 
$-$0.703                  & $-$0.614                  & $-$0.6821 \nl 
N2541, real+art. 1st epoch    & $-$0.869$\pm$0.003 (262) & 
$-$0.714$\pm$0.005 (175) & $-$0.618$\pm$0.004 (220) & 
$-$0.712$\pm$0.004 (210)\nl
N2541, real+art. 1st exposure & $-$0.880$\pm$0.005 (151) & 
$-$0.713$\pm$0.007 (111) &  \nodata\tablenotemark{2}                 & \nodata\tablenotemark{2} \nl
N2541, artificial frame       & $-$0.872$\pm$0.002 (524) & 
$-$0.683$\pm$0.001 (737) & $-$0.581$\pm$0.001 (558) & 
$-$0.682$\pm$0.002 (535)\nl
N1365, real+art. 1st epoch    & $-$0.891$\pm$0.005  (138)& 
$-$0.693$\pm$0.008 (80)  & $-$0.617$\pm$0.006 (156) & 
$-$0.708$\pm$0.006 (97)\nl
N1365, real+art. 1st exposure & $-$0.892$\pm$0.011  (34)   & 
$-$0.730$\pm$0.02  (27)   &  \nodata\tablenotemark{2}                 & \nodata\tablenotemark{2} \nl
N1365, artificial frame       & $-$0.869$\pm$0.002 (610) & 
$-$0.677$\pm$0.001 (726) & $-$0.579$\pm$0.001 (730) & 
$-$0.685$\pm$0.001 (662)\nl
\multicolumn{5}{l} {--F814W--}\\
LeoI values                   & 1.066                 & 
$-$0.740                  & $-$0.763                   & $-$0.7641 \nl
N2541, real+art. 1st epoch    & 1.081$\pm$0.003 (423) & 
$-$0.734$\pm$0.004 (219) & $-$0.755$\pm$0.003 (271)   & 
$-$0.773$\pm$0.003 (263)\nl
N2541, real+art. 1st exposure & 1.079$\pm$0.005 (136) & 
$-$0.757$\pm$0.008 (64)   &  \nodata\tablenotemark{2}                  & \nodata\tablenotemark{2} \nl
N2541, artificial frame       & 1.107$\pm$0.001 (558) & 
$-$0.739$\pm$0.001 (802) & $-$0.739$\pm$0.001 (724)  & 
$-$0.763$\pm$0.001 (712)\nl
N1365, real+art. 1st epoch    & 1.059$\pm$0.005 (210) & 
$-$0.766$\pm$0.008 (96)  & $-$0.754$\pm$0.004 (219)   & 
$-$0.780$\pm$0.006 (140)\nl
N1365, real+art. 1st exposure & 1.041$\pm$0.013  (30)   & 
$-$0.814$\pm$0.028  (18)   &  \nodata\tablenotemark{2}                  & \nodata\tablenotemark{2} \nl
N1365, artificial frame       & 1.105$\pm$0.001 (697) & 
$-$0.740$\pm$0.001 (784) & $-$0.735$\pm$0.001 (861)  & 
$-$0.763$\pm$0.001 (763)\nl
\enddata
\tablenotetext{1}{The aperture corrections are based on the number
of stars shown in parentheses.}
\tablenotetext{2}{DoPHOT was run on single exposures (to test for biases introduced by CR hits) only for the PC and WF2.}
\end{deluxetable}

\begin{deluxetable}{lrrrr}
\tablecolumns{5}
\tablewidth{0pc}
\scriptsize
\tablecaption{ALLFRAME Aperture Corrections\label{tbl-3}}
\tablehead{
\colhead{} &
\colhead{PC (mag)\tablenotemark{1}} &
\colhead{WF2 (mag)\tablenotemark{1}} &
\colhead{WF3 (mag)\tablenotemark{1}} &
\colhead{WF4 (mag)\tablenotemark{1}} \\
}
\startdata
\multicolumn{5}{l} {--F555W--}\\
N2541 1st exp. &  0.009$\pm$0.012 (21)  &  0.007$\pm$0.0085 (24) &  
0.040$\pm$0.011 (21) &   0.023$\pm$0.020 (13) \nl
N2541 2nd exp.& $-$0.023$\pm$0.011 (23)  & $-$0.012$\pm$0.0090 (23) 
&  0.021$\pm$0.013 (19) &  $-$0.009$\pm$0.023 (10) \nl
N1365 1st exp.& $-$0.186$\pm$0.043 (13)  & $-$0.046$\pm$0.025 (16)  & 
$-$0.011$\pm$0.035 (8)  &  $-$0.082$\pm$0.023 (18)\nl
N1365 2nd exp.& $-$0.166$\pm$0.031 (10)  & $-$0.036$\pm$0.026 (16)  & 
$-$0.001$\pm$0.036 (9)  &  $-$0.082$\pm$0.022 (18)\nl 
\multicolumn{5}{l} {--F814W--}\\
N2541 1st exp.&  0.013$\pm$0.012 (18)  &  0.047$\pm$0.0084 (23) &  
0.090$\pm$0.010 (22) &   0.069$\pm$0.014 (17) \nl
N2541 2nd exp.&  0.032$\pm$0.012 (16)  &  0.022$\pm$0.0083 (22) &  
0.037$\pm$0.010 (20) &   0.038$\pm$0.015 (15) \nl
N1365 1st exp.& $-$0.063$\pm$0.024 (9)   & $-$0.031$\pm$0.024 (13)  
&  0.023$\pm$0.035 (10) &   0.058$\pm$0.025 (12) \nl
N1365 2nd exp.& $-$0.342$\pm$0.111 (9)   & $-$0.039$\pm$0.028 (14)  & 
$-$0.008$\pm$0.031 (10) &   0.047$\pm$0.021 (13) \nl 
\enddata
\tablenotetext{1}{The aperture corrections are based on the number
of stars shown in parentheses.}
\end{deluxetable}

\begin{deluxetable}{lcccc}
\tablecolumns{5}
\tablewidth{0pc}
\scriptsize
\tablecaption{Control Test on the NGC 1365 Artificial Frames\label{tbl-3}}
\tablehead{
\colhead{} & 
\multicolumn{2}{c} {DoPHOT} &
\multicolumn{2}{c} {ALLFRAME}\\
\colhead{} & 
\colhead{$\Delta m \pm \sigma(\Delta m)$} &
\colhead{$\gamma \pm \sigma(\gamma)$} &
\colhead{$\Delta m \pm \sigma(\Delta m)$} &
\colhead{$\gamma \pm \sigma(\gamma)$} \\
\colhead{} & 
\colhead{(mag)} &
\colhead{(mag/mag)} &
\colhead{(mag)} &
\colhead{(mag/mag)} \\
}
\startdata
PC, F555W  & $-$0.20 $\pm$ 0.04 & $-$0.0010 $\pm$ 0.0010 & 0.00 $\pm$ 
0.03 & 0.0013 $\pm$ 0.0006 \nl
WF2, F555W & $-$0.04 $\pm$ 0.03 & $-$0.0020 $\pm$ 0.0008 & 0.01 $\pm$ 
0.03 & 0.0006 $\pm$ 0.0006 \nl
PC, F814W  & $-$0.08 $\pm$ 0.04 & $-$0.0018 $\pm$ 0.0009 & 0.00 $\pm$ 
0.03 & 0.0004 $\pm$ 0.0006  \nl
WF2, F814W & $-$0.08 $\pm$ 0.03 & $-$0.0009 $\pm$ 0.0007 & 0.00 $\pm$ 
0.03 & 0.0008 $\pm$ 0.0006  \nl
\enddata
\end{deluxetable}

\begin{deluxetable}{lrrrr}
\tablecolumns{5}
\tablewidth{0pc}
\scriptsize
\tablecaption{Photometric Test on the Real+Artificial Frames\label{tbl-3}}
\tablehead{
\colhead{} & 
\multicolumn{2}{c} {DoPHOT} &
\multicolumn{2}{c} {ALLFRAME}\\
\colhead{} & 
\colhead{$\Delta m \pm \sigma(\Delta m)$} &
\colhead{$\gamma \pm \sigma(\gamma)$} &
\colhead{$\Delta m \pm \sigma(\Delta m)$} &
\colhead{$\gamma \pm \sigma(\gamma)$} \\
\colhead{} & 
\colhead{(mag)} &
\colhead{(mag/mag)} &
\colhead{(mag)} &
\colhead{(mag/mag)} \\
}
\startdata
\multicolumn{5}{l}{--N1365--}\nl
PC, F555W  &  0.12 $\pm$ 0.05 & $-$0.0009 $\pm$ 0.0014 &  0.01 $\pm$ 
0.05 &  0.0007 $\pm$  0.0016\nl
WF2, F555W &  0.02 $\pm$ 0.05 &  0.0019 $\pm$ 0.0015 &  0.03 $\pm$ 
0.10 & $-$0.0022 $\pm$  0.0031\nl
WF3, F555W &  0.02 $\pm$ 0.04 &  0.0025 $\pm$ 0.0011 &  0.00 $\pm$ 
0.04 &  0.0012 $\pm$  0.0014\nl  
WF4, F555W &  $-$0.05 $\pm$ 0.05 &  0.0014 $\pm$ 0.0013 &  0.01 $\pm$ 
0.05 &  0.0026 $\pm$  0.0018\nl
PC, F814W  &  $-$0.02 $\pm$ 0.06 &  0.0013 $\pm$ 0.0018 &  0.01 $\pm$ 
0.04 &  0.0023 $\pm$  0.0016\nl 
WF2, F814W &  0.07 $\pm$ 0.06 &  0.0015 $\pm$ 0.0019 &  0.03 $\pm$ 
0.07 &  0.0009 $\pm$  0.0025\nl
WF3, F814W &  0.09 $\pm$ 0.04 &  0.0006 $\pm$ 0.0011 &  0.00 $\pm$ 
0.04 & $-$0.0010 $\pm$  0.0013 \nl
WF4, F814W &  0.10 $\pm$ 0.04 & $-$0.0015 $\pm$ 0.0013 &  0.00 $\pm$ 
0.04 &  0.0014 $\pm$  0.0015 \nl
\multicolumn{5}{l}{--N2541--}\nl
PC, F555W  &  0.02 $\pm$ 0.05 &  0.0005 $\pm$ 0.0013 &  0.09 $\pm$ 
0.05 & $-$0.0009 $\pm$  0.0013 \nl
WF2, F555W &  0.04 $\pm$ 0.04 &  0.0000 $\pm$ 0.0012 &  0.01 $\pm$ 
0.07 & $-$0.0053 $\pm$  0.0021 \nl
WF3, F555W &  0.07 $\pm$ 0.04 &  0.0017 $\pm$ 0.0011 &  0.01 $\pm$ 
0.05 & $-$0.0007 $\pm$  0.0014 \nl
WF4, F555W &  0.02 $\pm$ 0.05 & $-$0.0004 $\pm$ 0.0012 &    0.00 
$\pm$ 0.04 & $-$0.0034 $\pm$  0.0013 \nl
PC, F814W  &  0.00 $\pm$ 0.03 & $-$0.0002 $\pm$ 0.0009 &  0.00 $\pm$ 
0.03 & $-$0.0007 $\pm$  0.0010 \nl
WF2, F814W &  0.01 $\pm$ 0.04 &  0.0023 $\pm$ 0.0012 &  0.01 $\pm$ 
0.05 & $-$0.0022 $\pm$  0.0015 \nl
WF3, F814W &  0.08 $\pm$ 0.04 &  0.0019 $\pm$ 0.0010 &  0.00 $\pm$ 
0.03 & $-$0.0011 $\pm$  0.0011 \nl
WF4, F814W &  0.02 $\pm$ 0.03 &  0.0009 $\pm$ 0.0009 &    0.00 $\pm$ 
0.04 & $-$0.0002 $\pm$  0.0010 \nl
\enddata
\end{deluxetable}

\begin{deluxetable}{lcccc}
\tablecolumns{5}
\tablewidth{0pc}
\scriptsize
\tablecaption{Bias Affecting Stars in Single Epoch Photometry\label{tbl-3}}
\tablehead{
\colhead{Chip} &
\multicolumn{2}{c} {DoPHOT} &
\multicolumn{2}{c} {ALLFRAME}\\
\colhead{} & 
\colhead{$\epsilon_V$} &
\colhead{$\epsilon_I$} &
\colhead{$\epsilon_V$} &
\colhead{$\epsilon_I$} \\
\colhead{} & 
\colhead{(mag)} &
\colhead{(mag)} &
\colhead{(mag)} &
\colhead{(mag)} \\
}
\startdata
\multicolumn{5}{l}{--N1365--}\nl
PC  & 0.069  &   0.066 & 0.124  &   0.134\nl
WF2 & 0.138  &   0.179 & 0.207  &   0.191\nl
WF3 & 0.074  &   0.072 & 0.119  &   0.124\nl
WF4 & 0.082  &   0.107 & 0.158  &   0.166\nl
\multicolumn{5}{l}{--N2541--}\nl
PC  & 0.040  &   0.041 & 0.034  &   0.080\nl
WF2 & 0.088  &   0.118 & 0.121  &   0.145\nl
WF3 & 0.039  &   0.043 & 0.051  &   0.098\nl
WF4 & 0.041  &   0.037 & 0.049  &   0.068\nl
\enddata
\end{deluxetable}

\begin{deluxetable}{lcccc}
\tablecolumns{5}
\tablewidth{0pc}
\scriptsize
\tablecaption{Bias Affecting the Cepheid Apparent Distance Moduli\label{tbl-3}}
\tablehead{
\colhead{Chip} &
\multicolumn{2}{c} {DoPHOT} &
\multicolumn{2}{c} {ALLFRAME}\\
\colhead{} & 
\colhead{$\epsilon_V$} &
\colhead{$\epsilon_I$} &
\colhead{$\epsilon_V$} &
\colhead{$\epsilon_I$} \\
\colhead{} & 
\colhead{(mag)} &
\colhead{(mag)} &
\colhead{(mag)} &
\colhead{(mag)} \\
}
\startdata
\multicolumn{5}{l}{--N1365--}\nl
PC  & 0.040  &   0.037 &  0.026  &   0.021\nl  
WF2 & 0.086  &   0.075 &  0.036  &   0.056\nl  
WF3 & 0.033  &   0.041 &  0.021  &   0.018\nl  
WF4 & 0.055  &   0.056 &  0.026  &   0.031\nl  
\multicolumn{5}{l}{--N2541--}\nl
PC  & 0.027  &   0.027 &  0.000  &   0.017\nl  
WF2 & 0.063  &   0.067 &  0.037  &   0.028\nl  
WF3 & 0.023  &   0.011 &  0.009  &   0.021\nl  
WF4 & 0.031  &   0.018 &  0.022  &   0.013\nl  
\enddata
\end{deluxetable}

\clearpage

\begin{figure}
\epsscale{0.6}
\caption{Top: A CR-free image of the NGC 1365 field obtained
by combining the eight F555W exposures in Table 1. Bottom: the
artificial star frame added to the NGC 1365 field. The two images are
displayed with the same grey-scale, to preserve the brightness ratio of
the real and artificial stars.}
\end{figure}

\clearpage

\begin{figure}
\epsscale{0.6}
\caption{As for Figure 1, but for the NGC 2541 field.}
\end{figure}

\clearpage

\begin{figure}
\epsscale{1.0}
\plotone{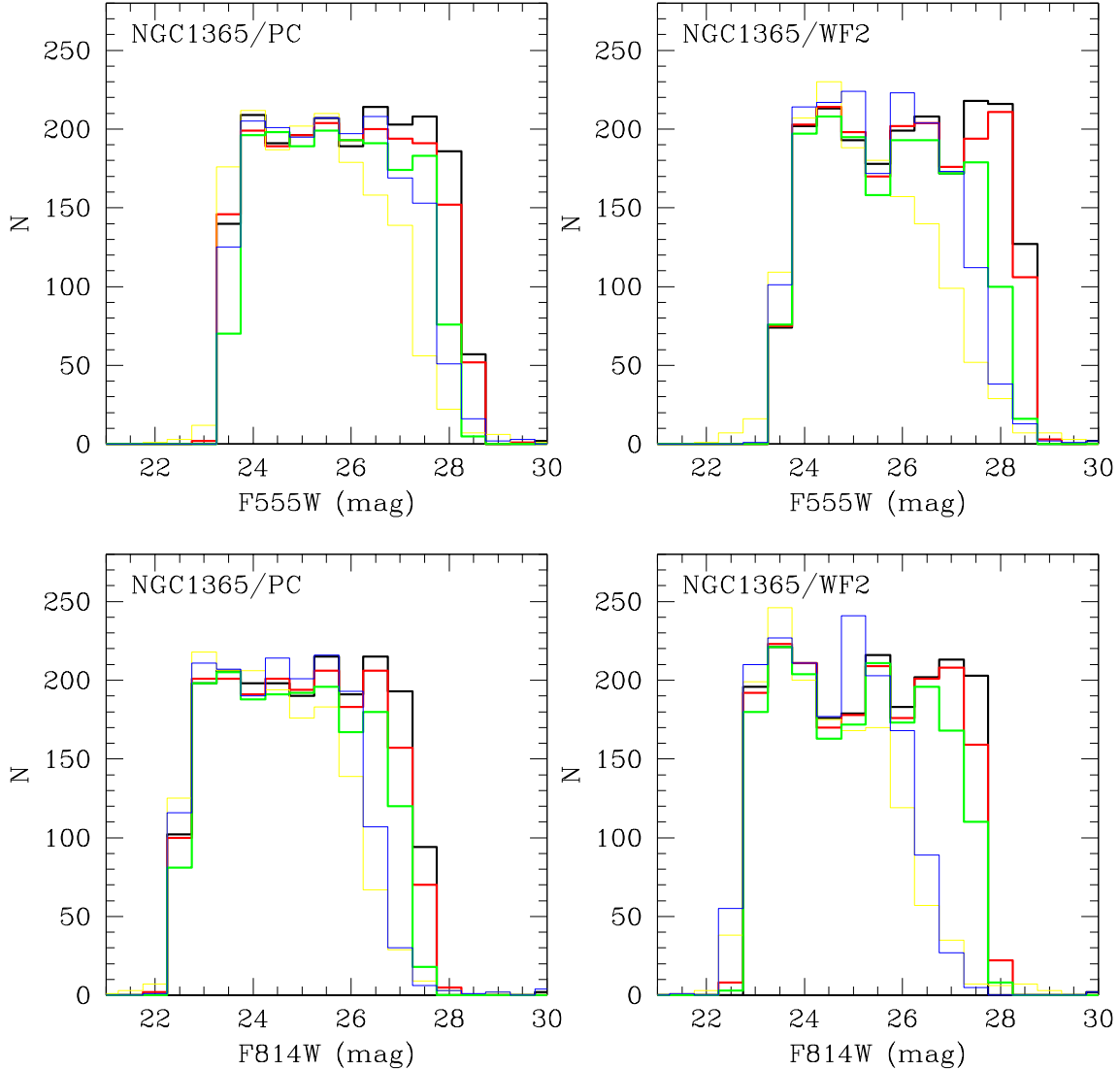}\\
\caption{Luminosity function (LF) of the input and recovered stars in
the PC and WF2 NGC 1365 fields. The input LF is in black. The LF's recovered
from the frames containing only the artificial stars by ALLFRAME and
DOPHOT are shown by the thick red and green lines respectively. The
thin yellow and blue lines are the LF's for the artificial stars
recovered by ALLFRAME and DoPHOT respectively in the artificial+real
star frames. The photometry for the first exposure, U2S71201T, was
used for ALLFRAME, and for the first epoch (U2S71201T and U2S71201T
combined) for DoPHOT).}
\end{figure}

\begin{figure}
\plotone{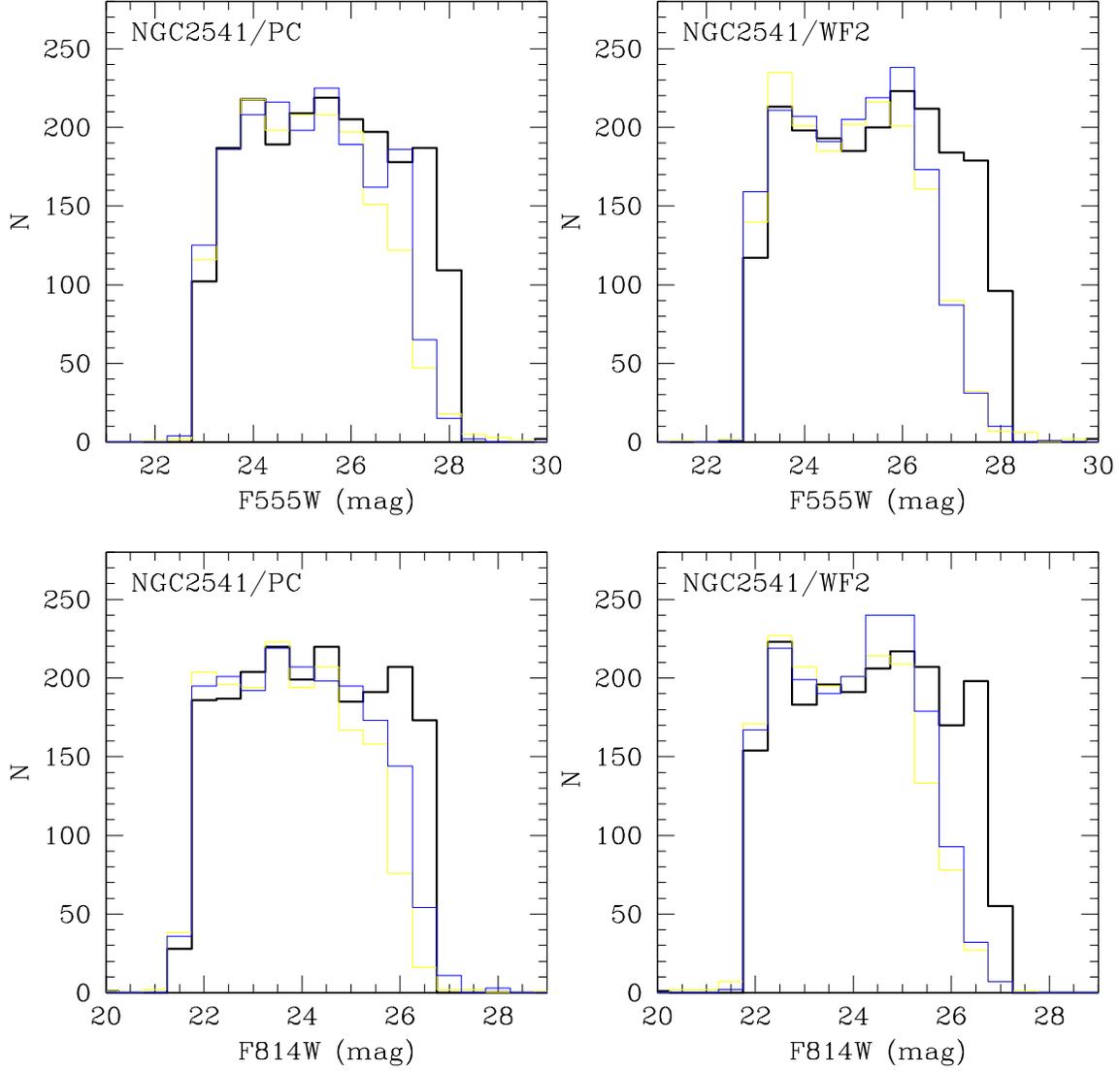}\\
\caption{As Figure 3, but for the NGC 2541 field. The luminosity
functions recovered from the artificial star frames are not shown. The
photometry used for the real+artificial star frames is from US72901T
for ALLFRAME, and the first epoch (US72901T and US72902T combined) 
for DoPHOT.}
\end{figure}

\begin{figure}
\plotone{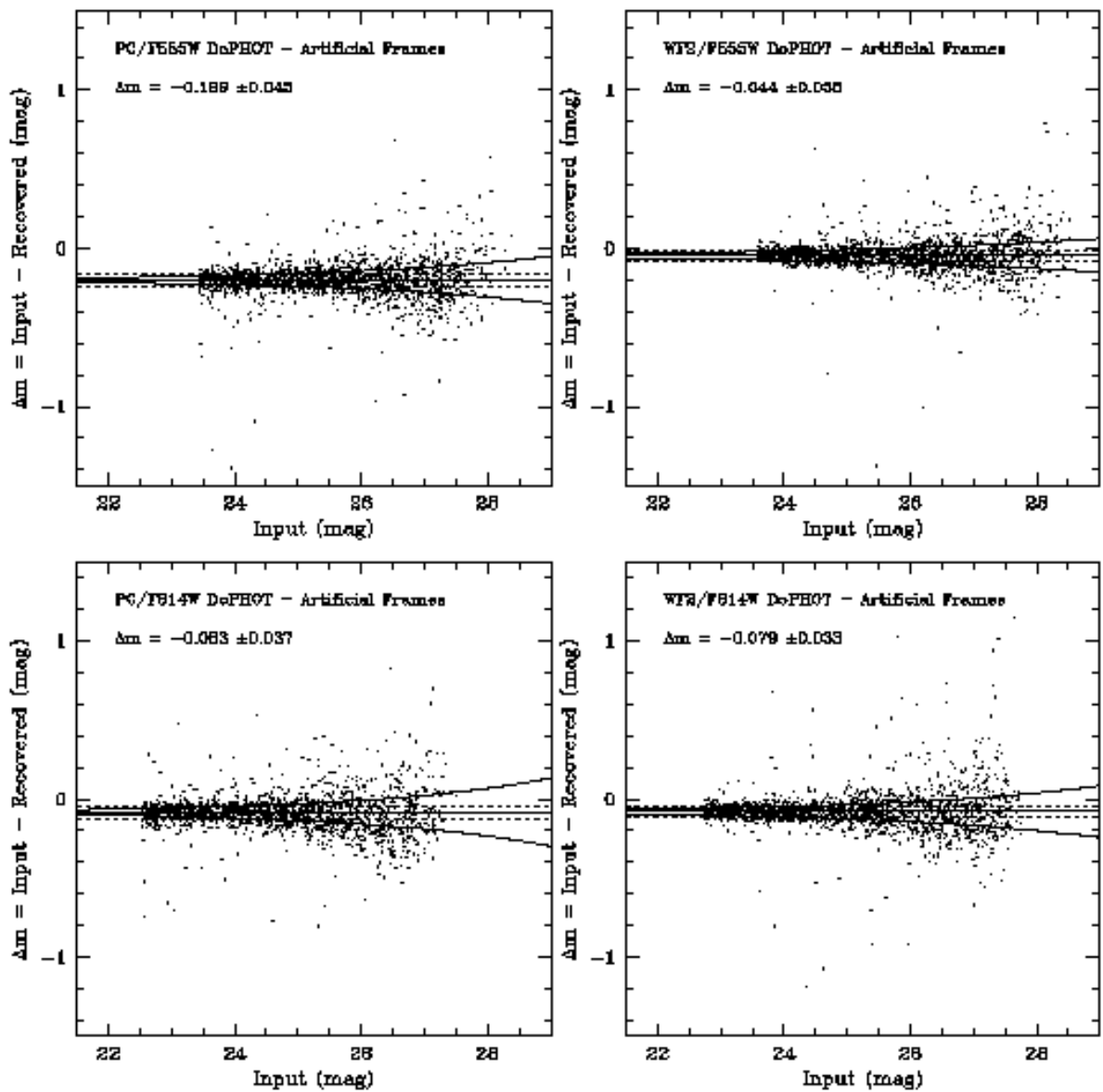}\\
\caption{Difference between the input magnitudes and the magnitudes
recovered by DoPHOT from the artificial star frames. The exponential
curves represent the typical error reported by DoPHOT for a star
of the magnitude shown in the abscissa. The solid and dotted straight
lines  are the weighted mean and 1$\sigma$ deviations calculated as
explained in the text.}
\end{figure}

\begin{figure}
\plotone{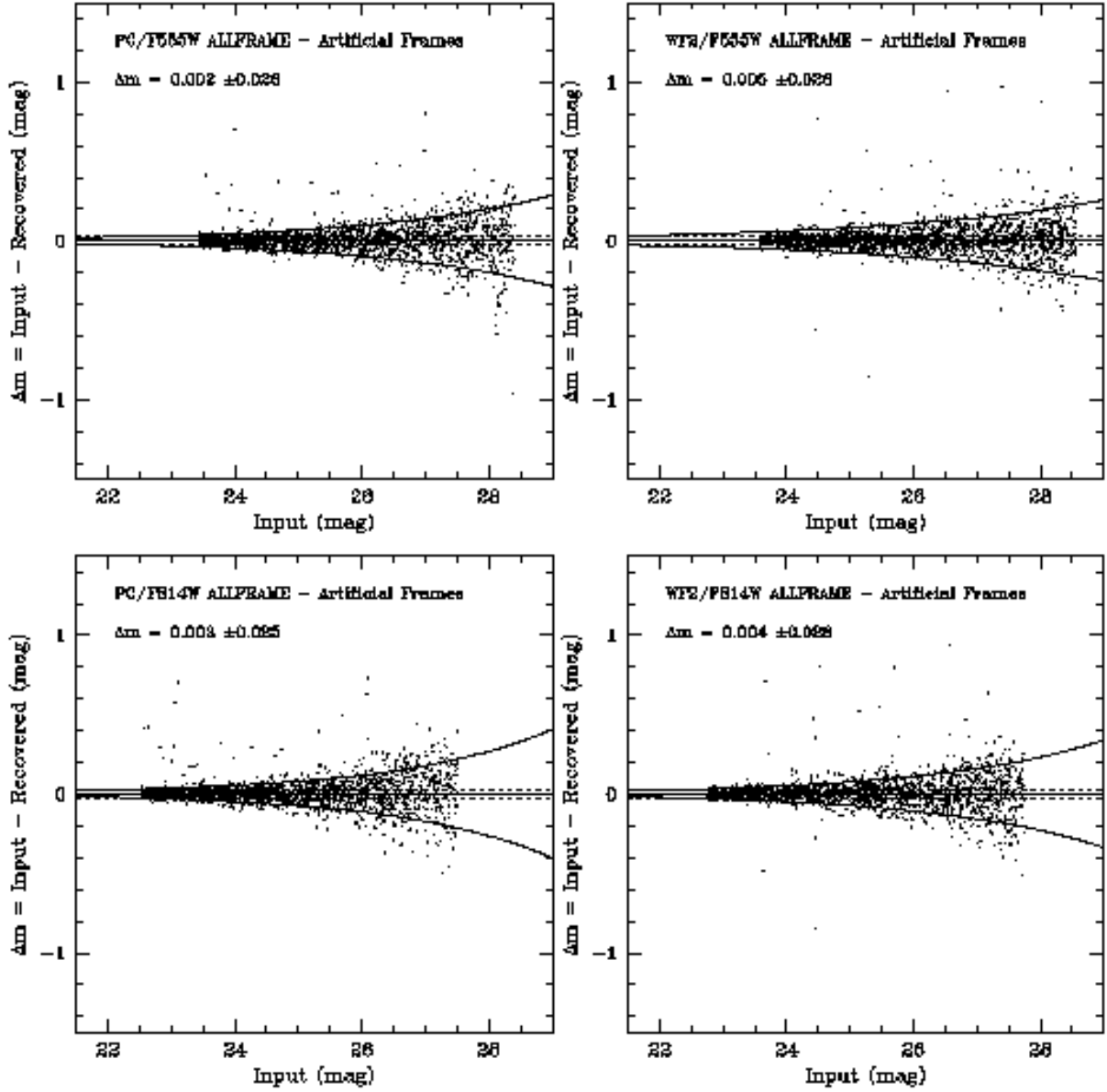}\\
\caption{As Figure 5, but for the ALLFRAME reduction.}
\end{figure}

\begin{figure}
\plotone{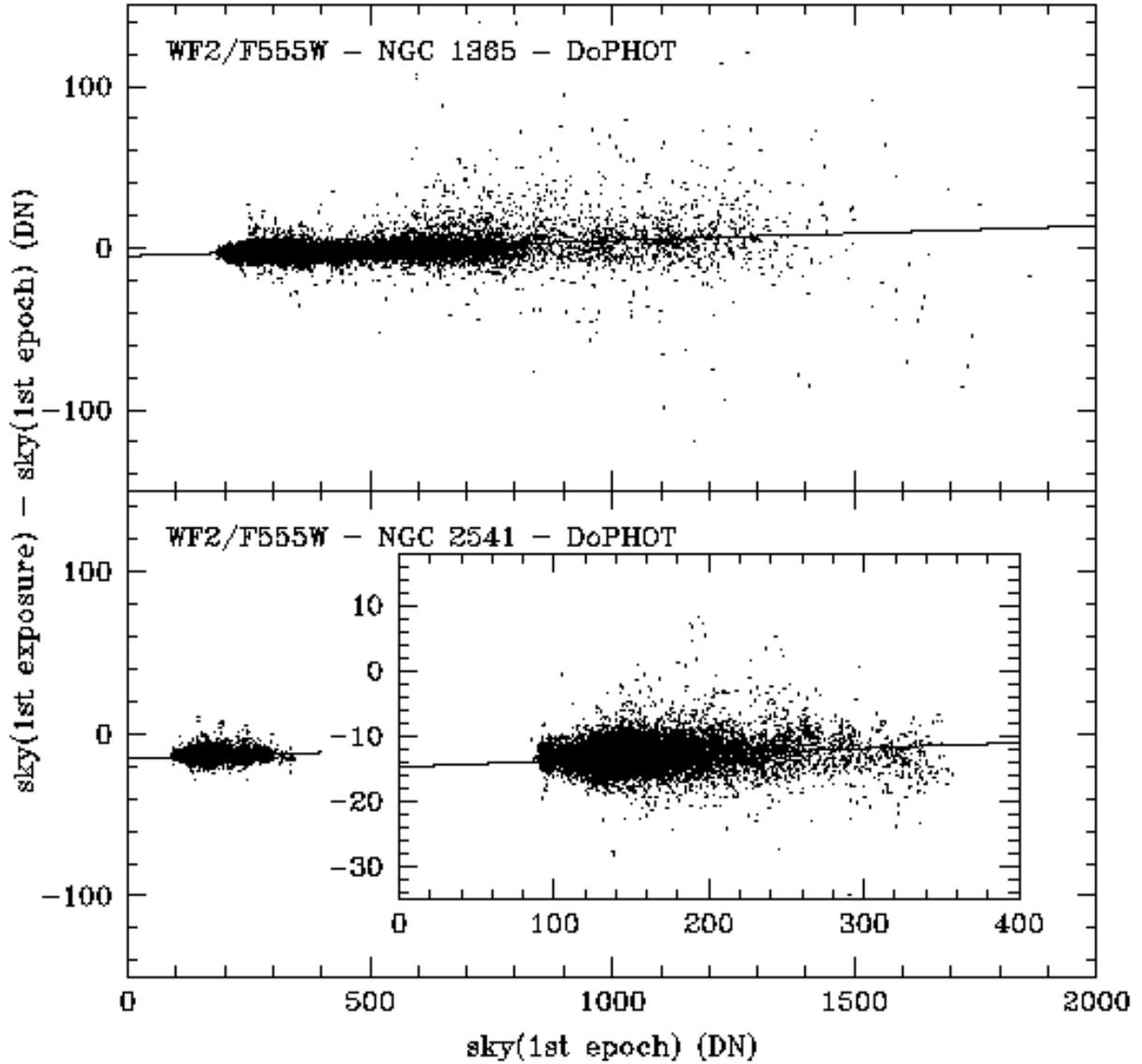}\\
\caption{Difference in the sky values measured by DoPHOT for the
stars in the first exposure and first epoch (first and second
exposure combined) of NGC 1365 (top) and NGC 2541 (bottom). The solid
line is a least-squares fit to the data.}
\end{figure}

\begin{figure}
\plotone{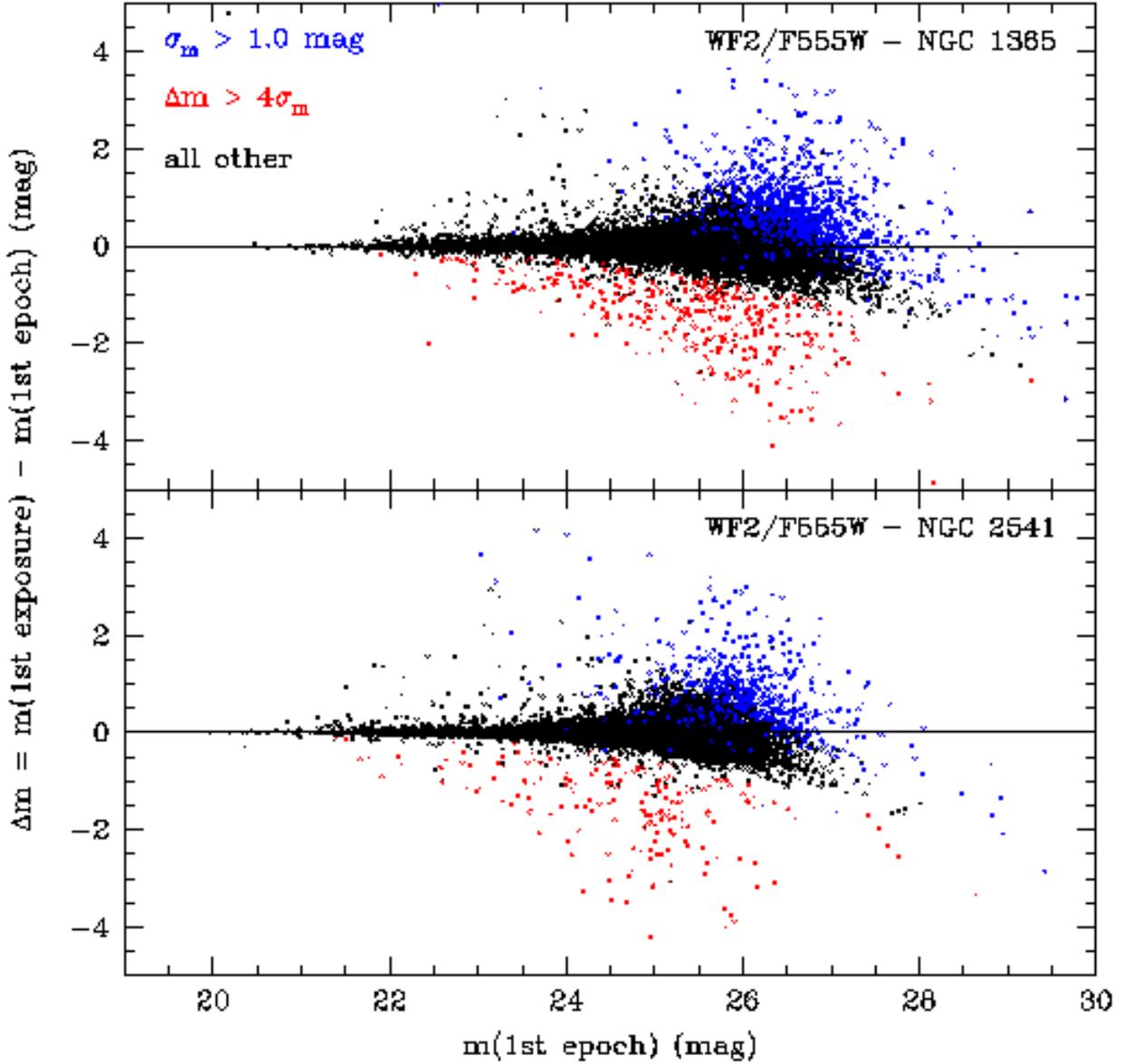}\\
\caption{Difference between the magnitudes measured by DoPHOT for the
stars in the first exposure and first epoch. Points for which this
difference is larger than four times the photometric error measured in
the first epoch are shown in red, points for which the photometric
error is larger than 1 mag are in blue.}
\end{figure}

\begin{figure}
\plotone{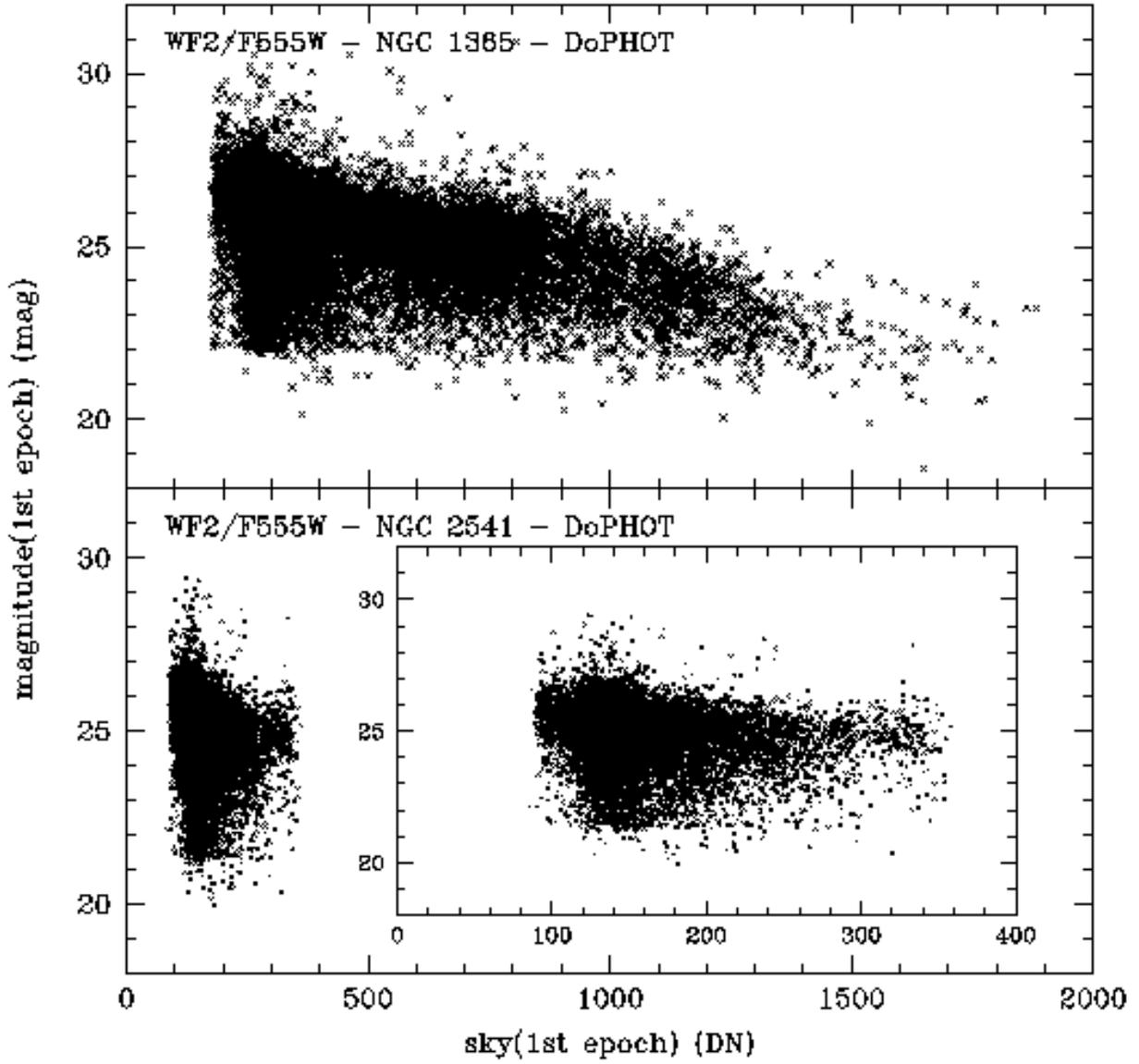}\\
\caption{Completeness as a function of background brightness (in total
number of counts) for the first epoch of NGC 2541 and NGC 1365.}
\end{figure}

\begin{figure}
\plotone{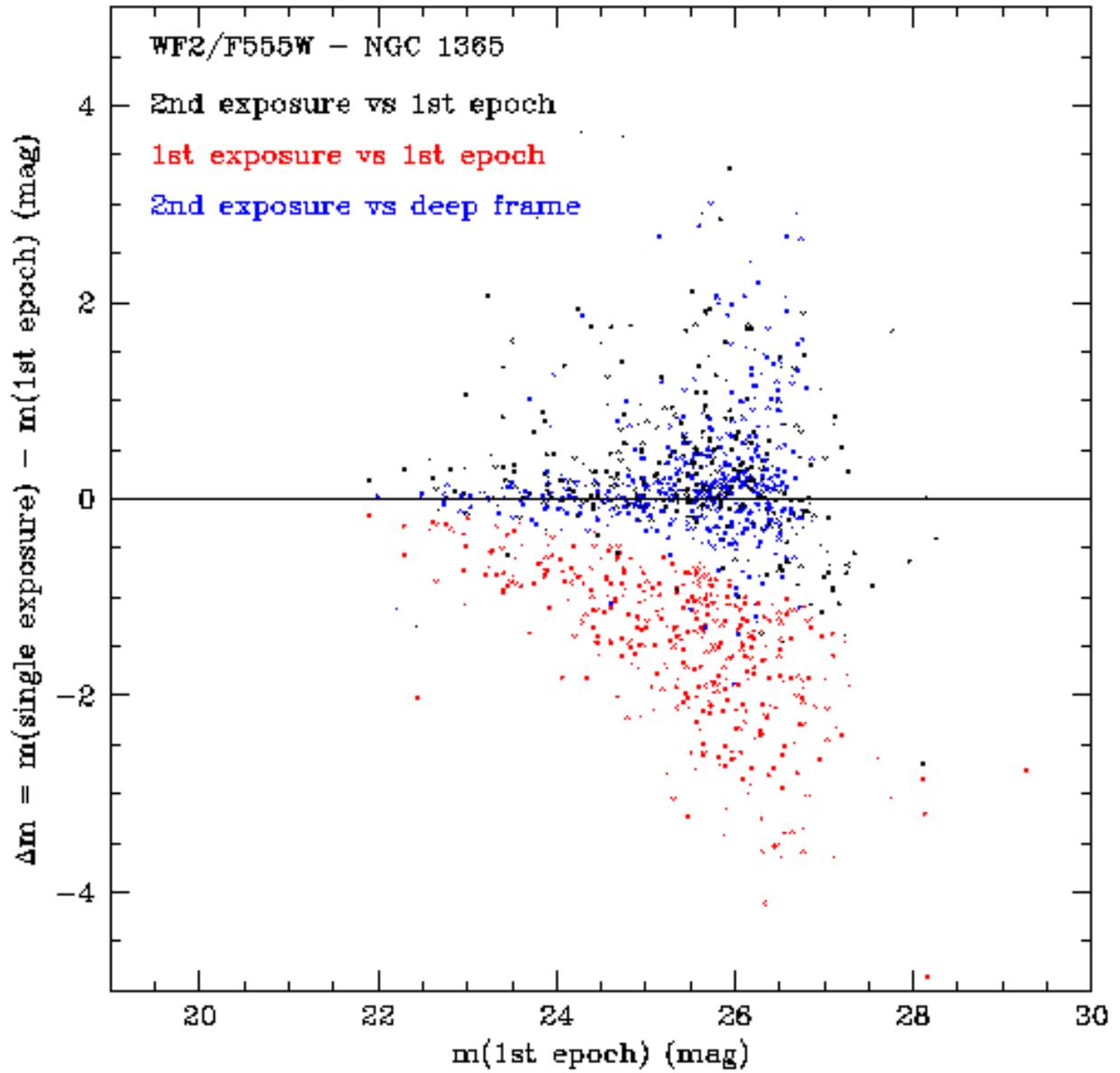}\\
\caption{The red points are the same as in Figure 8. These stars have
been singled out and their magnitude measured in the second exposure
of the CR-split pair for the first epoch is compared to the one
measured in the first epoch (black points) and in the deep frames (all
epochs combined, blue points).}
\end{figure}

\begin{figure}
\plotone{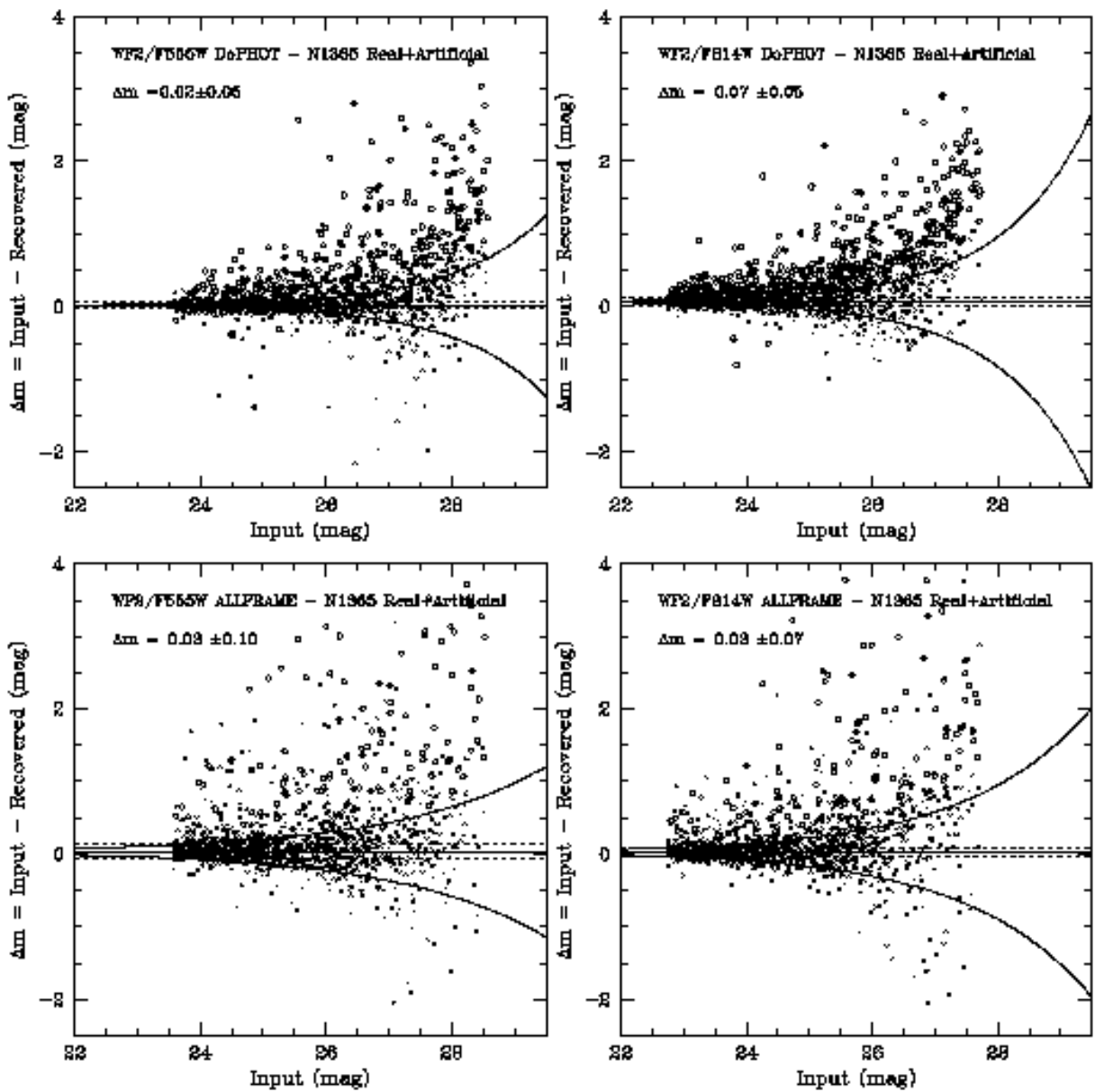}\\
\caption{Comparison between the input magnitudes and the magnitudes
derived by DoPHOT (top panels) and ALLFRAME (lower panels) from the
real+artificial WF2 NGC 1365 frames. The exponential
curves represent the typical error for stars whose
magnitude is shown in the abscissa. Crosses corresponds to stars for
which the difference is smaller than three times the
measured magnitudes error $\sigma$, open circles to
stars with difference $> 3\sigma$ mag.}
\end{figure}

\begin {figure}
\plotone{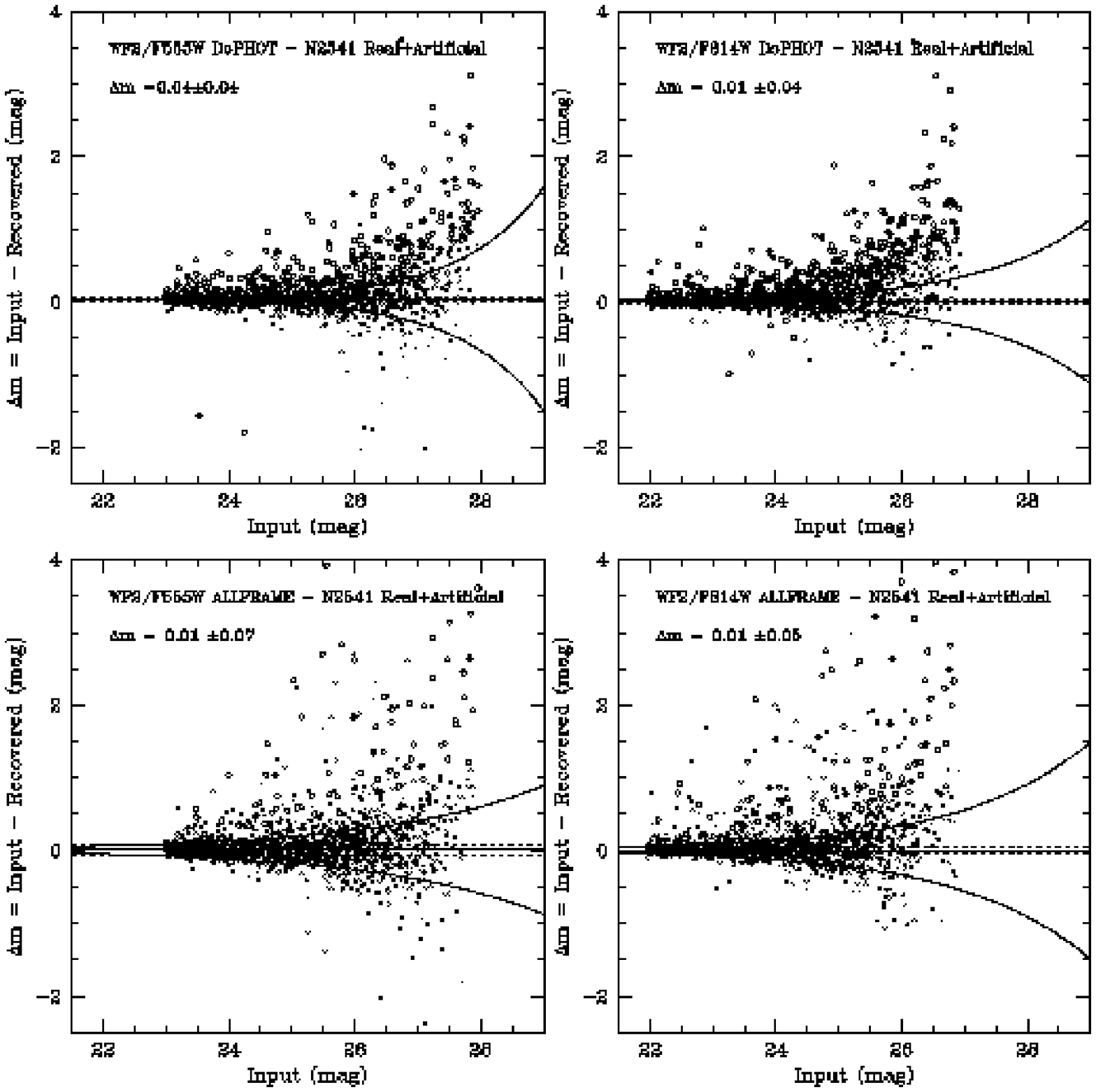}\\
\caption{As Figure 11, but for NGC 2541.}
\end{figure}

\clearpage

\begin {figure}
\plotone{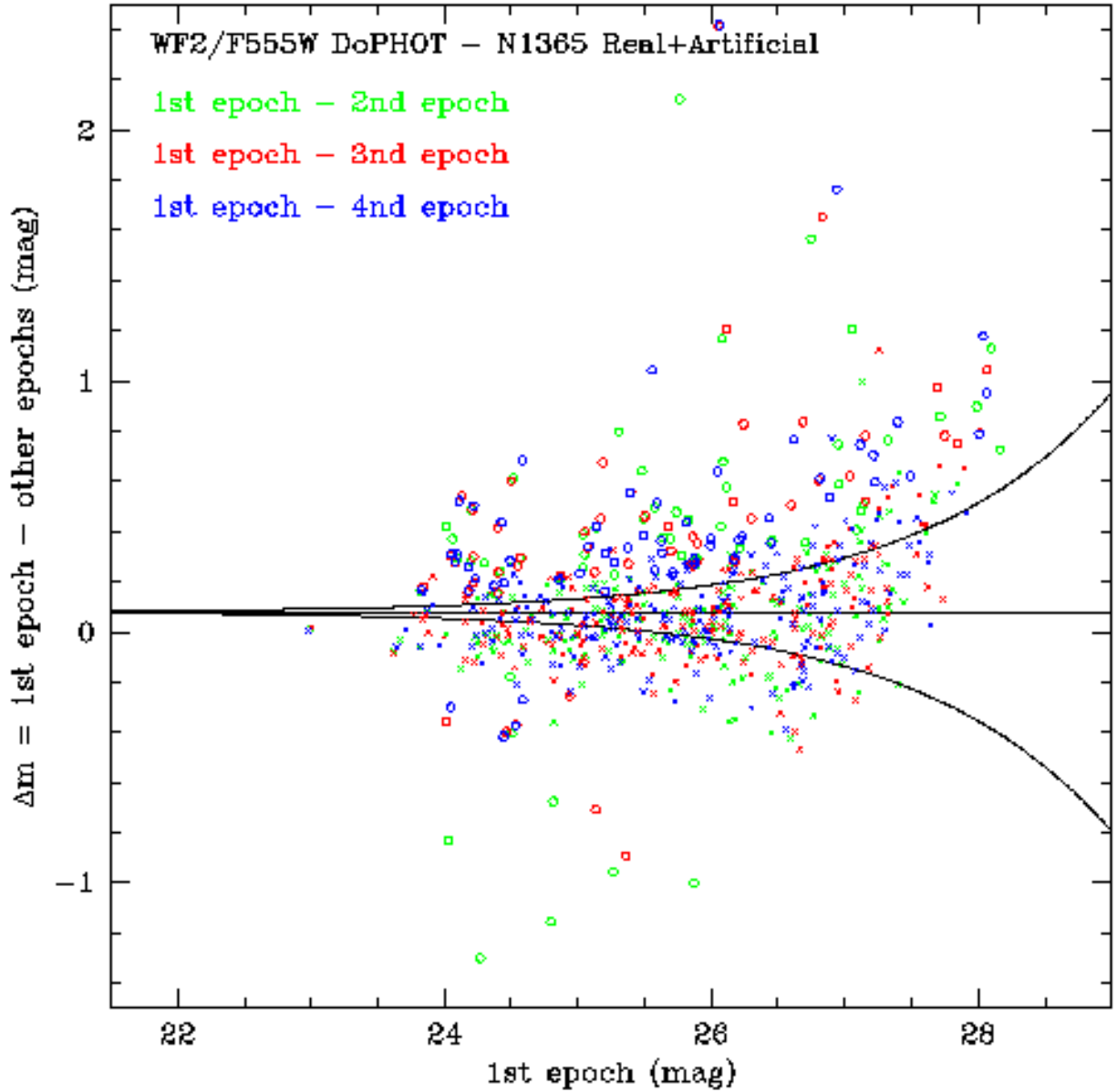}\\
\caption{Comparison between the F814W magnitudes derived by DoPHOT for
the four different epochs of the real+artificial NGC 1365 frames. The
y axis plots the difference with respect to epoch 1 for epochs 2, 3
and 4, as shown by the color codes at the top of the figure. The
crosses correspond to stars which are measured consistently too
bright in all epochs due to confusion noise. The circles are stars
which are measured too bright only in the first epoch because of a
transient phenomenon, such as an unidentified cosmic ray event, but
are measured correctly in all other epochs.}
\end{figure}

\begin{figure}
\plotone{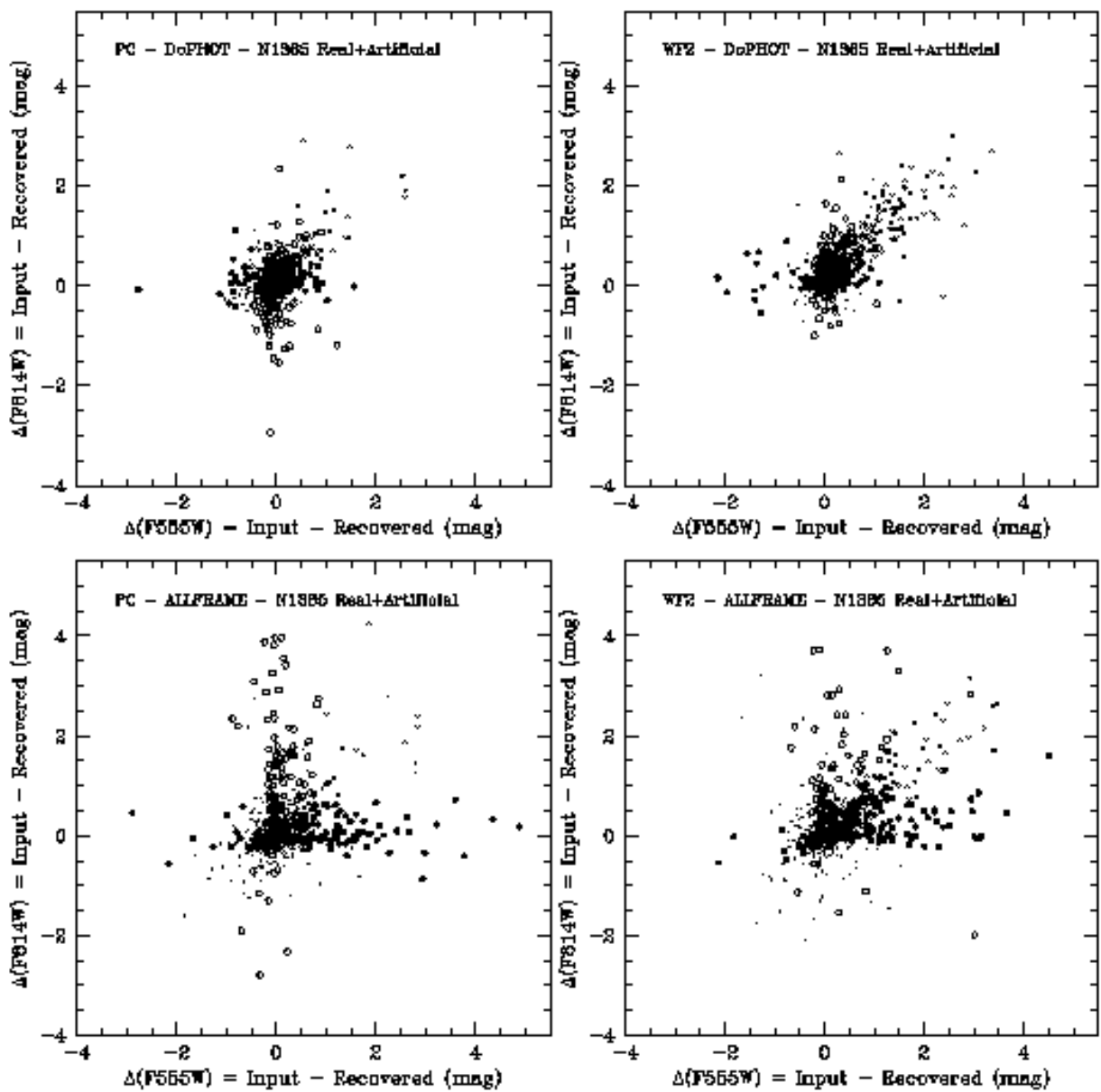}\\
\caption{Correlation between the input and recovered magnitudes in the
F555W and F814W passbands. Small points are for stars which deviate by
less than three times the reported error $\sigma$ in both passbands, filled
and open circles are points that deviate by more than 3$\sigma$ in
F555W or F814W respectively, and crosses are for points deviating by
more than 3$\sigma$ in both passbands. Only the PC and WF2 are shown, DoPHOT 
photometry is plotted in the upper two panels, ALLFRAME in the lower two.}
\end{figure}

\begin{figure}
\plotone{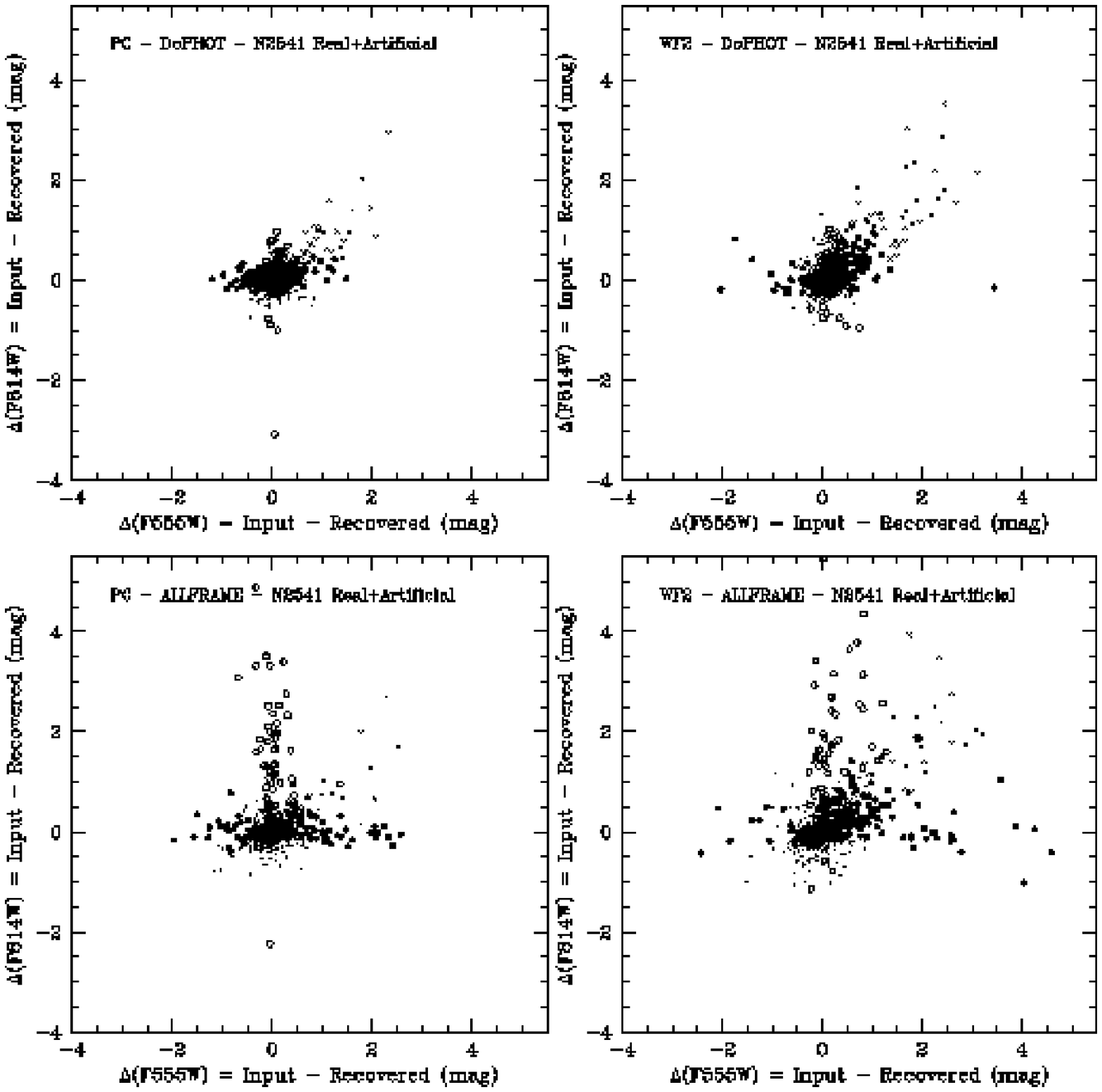}\\
\caption{As Figure 14, but for NGC 2541.}
\end{figure}

\begin{figure}
\plotone{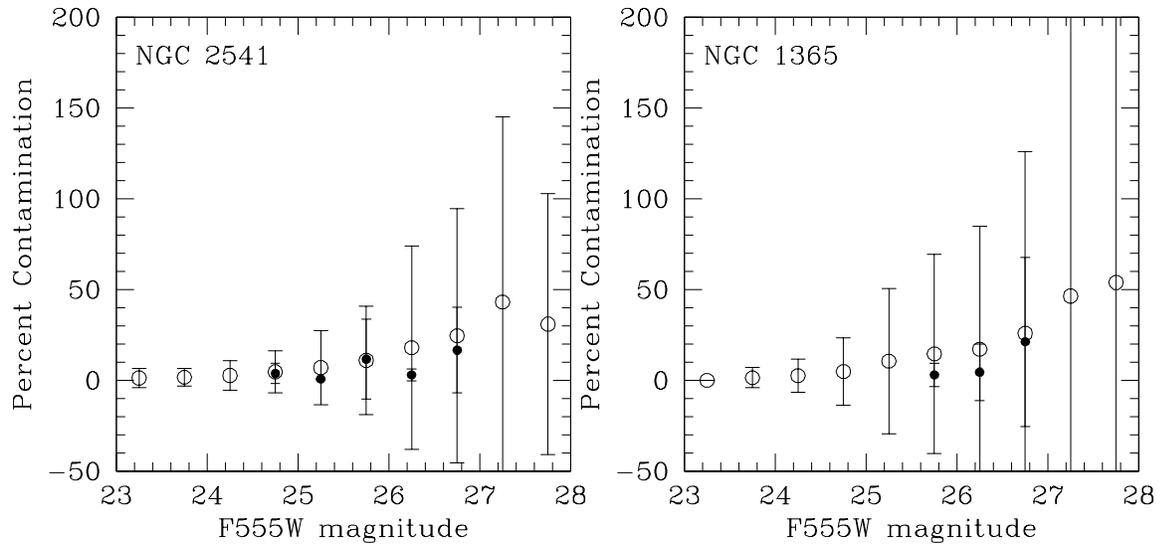}\\
\caption{The mean and rms contamination affecting the
sample of Cepheids (solid circles and smaller errorbars) and
artificial stars (open circles and larger errorbars) as a function of
F555W magnitudes (see text for further details).}
\end{figure}

\begin{figure}
\plotone{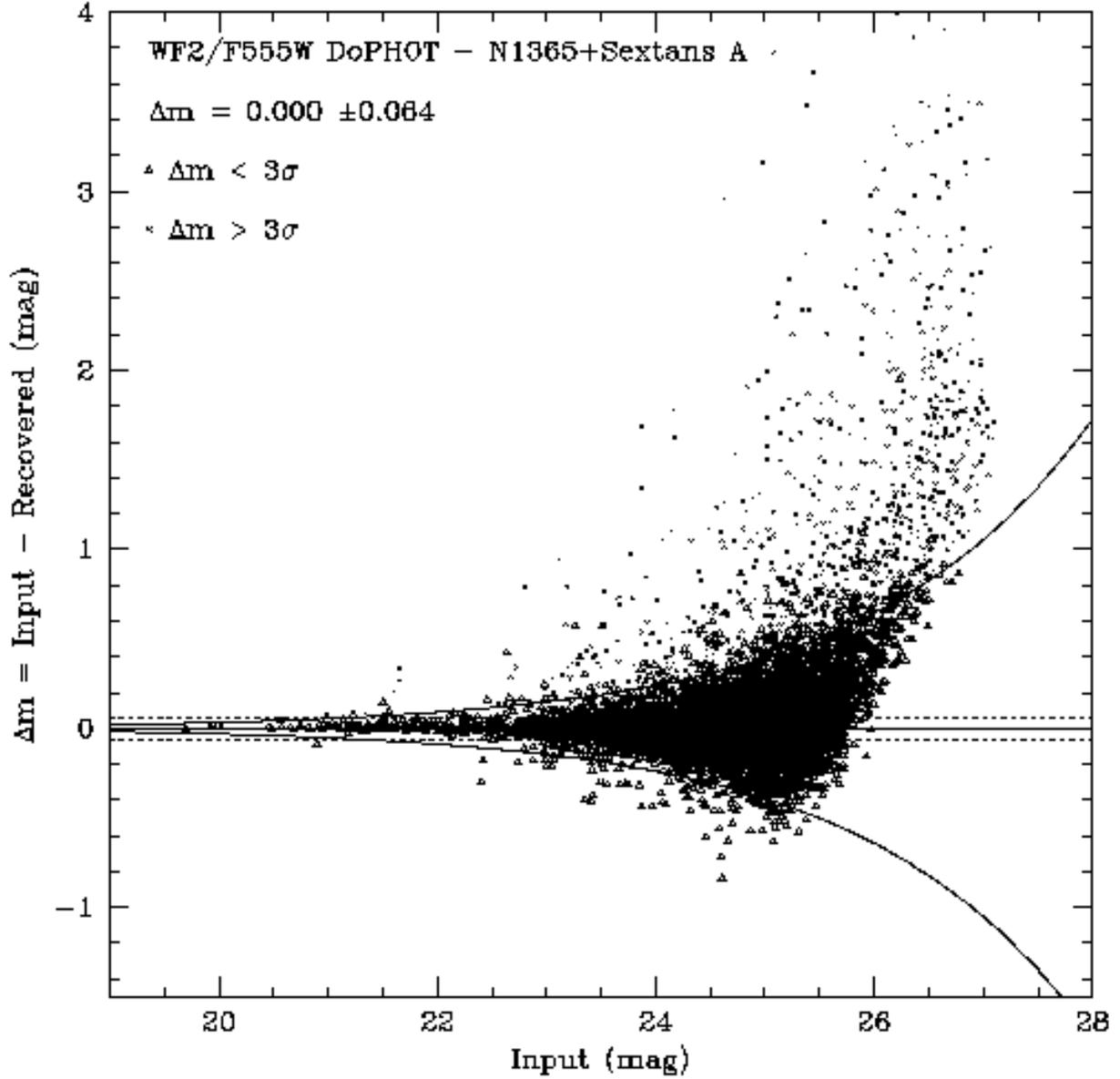}\\
\caption{Photometric Zero point test conducted using an uncrowded
Sextans A  F555W/WF2 field. Input magnitudes are DoPHOT measured
magnitudes in the Sextans A frame, recovered magnitudes are DoPHOT
magnitudes measured when the Sextans A field is added to the F555W/WF2
first epoch of NGC 1365.}
\end{figure}

\begin{figure}
\plotone{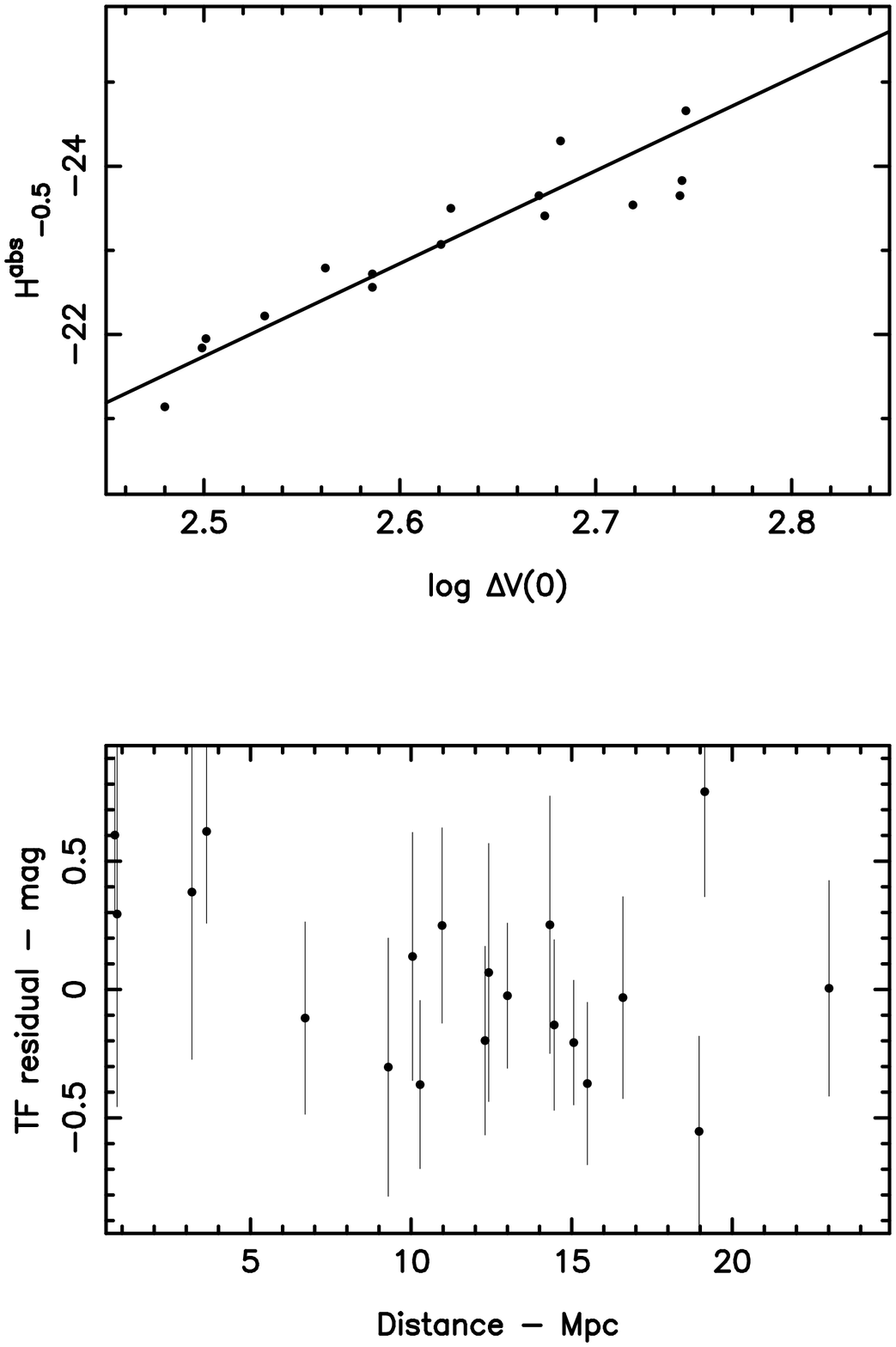}\\
\caption{The upper panel plots the H-band Tully-Fisher relation  from
Sakai \etal (2000), while in the lower panel the H-band residuals are
plotted against distance.}
\end{figure}


\begin{references}

Ferrarese, L., \etal 1996, ApJ, 464, 568

Ferrarese, L., \etal 1998, ApJ, 507, 655

Gibson, B. K., Maloney, P. R., \& Sakai, S. 2000, ApJL, submitted

Hill, R., \etal 1998, ApJ, 496, 648

Holtzmann, J. A. \etal 1995, PASP, 107, 1065

Kennicutt, R. C., Freedman, W. L.,  \&  Mould, J. R. 1995, AJ, 110, 1476

Mochejska, B. J., Macri, L. M., Sasselov, D. D., \& Stanek, K. Z. 1999, astro-ph/9908293 

Saha, A., Labhardt, L., Schwengeler, H., Macchetto, F. D., Panagia, N., Sandage, A. \&  Tammann, G. A. 1994, ApJ, 425, 14

Saha, A., Sandage, A. \&  Tammann, G. A., Labhardt, L., Macchetto, F. D., Panagia, N. 1997, ApJ, 486, 1

Saha, A., Labhardt, L., \& Prosser, C. 2000, PASP, in press

Sakai, S., \etal 2000, ApJ, in press (astro-ph/9909269)

Sandage, A., \etal 1992, ApJL, 401, 7

Schechter, P. L., Mateo, M., \& Saha, A. 1993, PASP, 105, 1342

Silbermann, N. A. \etal 1999, ApJ, 515, 1

Stanek, K. Z., \& Udalski, A. 1999, astro-ph/9909346

Stetson, P. B. 1990, PASP, 102, 932

Stetson, P. B. 1994, PASP, 106, 250

Stetson, P. B. 1998, PASP, 110, 1448

Whitmore, B., \& Heyer, I. 1997, Instrument Science Report WFPC2 97-08 (Baltimore: STScI)

\end{references}
\end{document}